\documentclass[aps,prd,twocolumn,showpacs,groupedaddress]{revtex4-1}

\usepackage[dvipdfmx]{graphicx}  
\usepackage{dcolumn}   
\usepackage{bm}        
\usepackage{amssymb}   
\usepackage{amsmath}	
\usepackage[dvipdfmx]{color}	
\usepackage{ulem}  
\newcommand{\eref}[1]{(\ref{#1})}

\newcommand{\dev}[0]{\mathrm d}
\newcommand{\bibun}[3]{\frac{\dev^{#3}{#1}}{\dev{#2}^{#3}}}
\newcommand{\henbibun}[3]{\frac{\partial^{#3}{#1}}{\partial {#2}^{#3}}}
\newcommand{\sekibun}[3]{\int_{#1}^{#2}\dev{#3}}
\newcommand{\im}[0]{i}
\newcommand{\ex}[1]{{e^{#1}}}

\numberwithin{equation}{section}
\renewcommand\theequation{\arabic{section}.\arabic{equation}}

\newcommand{\mass}[0]{m}
\newcommand{\pusai}[1]{\psi}

\renewcommand{\Re}[0]{{\rm Re}}
\renewcommand{\Im}[0]{{\rm Im}}

\begin{document}


\title{Gravitational field of a Schwarzschild black hole and a rotating mass ring}


\author{Yasumichi Sano and Hideyuki Tagoshi}
\affiliation{Department of Earth and Space Science, Graduate School of Science, Osaka University, Toyonaka, Osaka 560-0043}


\date{\today}

\begin{abstract}
The linear perturbation of the Kerr black hole has been discussed by using the 
Newman{--}Penrose
and the perturbed Weyl scalars, $\psi_0$ and $\psi_4$ can be obtained 
from the Teukolsky equation. 
In order to obtain the other Weyl scalars and the perturbed metric, 
a formalism was proposed by Chrzanowski and by Cohen and Kegeles (CCK)
to construct  these quantities in a radiation gauge 
via the Hertz potential.
As a simple example of the construction of the perturbed gravitational field with this formalism, 
we consider the gravitational field produced by 
a rotating circular ring around a Schwarzschild black hole. 
In the CCK method, the metric is constructed 
in a radiation gauge
via the Hertz potential, which is obtained from the solution of the Teukolsky equation. 
Since the solutions $\psi_0$ and $\psi_4$ of the Teukolsky equations are spin-2 quantities, 
the Hertz potential is determined up to its monopole and dipole modes.
Without these lower modes, the constructed metric and Newman--Penrose Weyl scalars have 
unphysical jumps on the spherical surface at the radius of the ring. 
We find that the jumps of the imaginary parts of the Weyl scalars are cancelled 
when we add the angular momentum perturbation to the Hertz potential. 
Finally, by adding the mass perturbation and choosing the parameters 
which are related to the gauge freedom, 
we obtain the perturbed gravitational field 
which is smooth except on the equatorial plane outside the ring.
We discuss the implication of these results to the problem of the computation of the gravitational 
self-force to the point particles in a radiation gauge.

\end{abstract}


\maketitle

\section{Introduction}
The black hole perturbation theory has been a powerful tool 
to investigate the stability of the 
black hole, the quasi-normal modes, 
and the gravitational waves produced by matters such like 
compact starts orbiting around the hole, and so on. 
For the Schwarzschild case, the first order metric perturbation is described 
by the Regge--Wheeler--Zerilli formalism \cite{RW1957, Z1970},
which relies on the spherical symmetry of the black hole space-time. 
The Regge--Wheeler and the Zerilli equation are the single, decoupled equation for 
the odd and even {parity} modes, respectively, and the master equations are reduced to 
radial ordinary differential equations by using the Fourier-harmonic expansion. 
On the other hand, for the Kerr case, it is well-known that there is no such a formalism 
for the metric perturbation. 
Instead, the perturbation of the Weyl scalars, $\psi_0$ and $\psi_4$, are described 
by the Teukolksy equation with the spin-weight $s=\pm 2$. 
One method to compute the metric perturbation of Kerr space-time is to solve the coupled 
partial differential equations numerically. 
The other method is to construct the metric perturbation from 
the perturbation of $\psi_0$ and $\psi_4$ obtained from the Teukolsky equation. 
Such a method was {proposed} first 
by Chrzanowski \cite{c} and Cohen and Kegeles \cite{kc, ck}
(See also \cite{wald1978, Stewart1979}),
{and thus} is called the CCK formalism. 
In this method, a radiation gauge is used to calculate the metric perturbation. 
After these works, 
however, there were very little development of the CCK formalism for a long time. 

New {developments were started} about a decade ago
by  Lousto and Whiting \cite{LoustoWhiting} and Ori \cite{ori03}.
These were motivated by the necessity to compute 
the gravitational self-force on the point particle orbiting around a Kerr black hole. 
Such situations are called EMRI (extreme mass ratio inspiral), and 
are one of the most important sources 
{of the gravitational wave} for the future space laser interferometers
such as eLISA \cite{eLISA}, DECIGO \cite{SNK, DECIGO} and BBO \cite{BBO}. 

A first explicit computation of the metric perturbation by using the CCK formalism 
was done by Yunes and Gonz$\acute{\rm a}$lez \cite{YunesGonzalez}
in which the vacuum perturbation was considered. 
Keidl, Friedman, and Wiseman \cite{kfw} were the first to find 
the explicit metric perturbation {produced} by a point particle, using the CCK formalism. 
They considered a system which consists of a Schwarzschild black hole and a static point mass, as a toy model. 
The metric perturbation is obtained straightforwardly for the multipole modes of $l\geq 2$. 
They obtained lower modes of $l=0, 1$ by considering the regularity of the metric.
A singularity, however, remained along a radial line which connect
the position of the particle and either the infinity or the black hole horizon. 
The presence of the singularity 
was previously discussed by Wald \cite{wald1973} and by Barack and Ori \cite{BarackOri2001}.

Keidl, Shah, Friedman, Kim and Price \cite{ksf+10, skf+11, sfk12}
{further developed}
the formalism to calculate the self-force by using the CCK formalism. 
In \cite{sfk12}, they reported the numerical corrections of gauge invariants 
of a particle in circular orbit {around} a Kerr black hole. 
For the calculation of the gravitational self{-}force
on the particle, 
it is important to complete the metric perturbation by adding the lower modes in an appropriate gauge.
The $l\geq 2$ modes are calculated in a radiation gauge, 
and the effects of lower modes are added in, what they call, the Kerr gauge. 

Recently, Pound, Merlin, and Barack \cite{pmb} discussed prescriptions for calculating the self-force 
from completed metric perturbations. 
With this prescription, 
once we obtain the metric perturbation which is constructed using {a} radiation gauge 
and completed 
with lower modes appropriately, it is possible to transform its gauge into a local Lorenz gauge. 
The regularized self-force can then be calculated by using the standard mode-sum method.

In this paper, we consider the metric perturbation of a rotating circular mass ring around a Schwarzschild black hole,
in order to understand the problems {in constructing}
the metric perturbation by using the CCK formalism. 
Especially, we discuss the problem of the completion of the metric perturbation with lower multipole modes. 
Of course, this is a first step toward the calculation of the metric perturbation produced by a orbiting particle.
But this problem is simpler than that of an orbiting particle, 
since the ring is circular and rotates with a constant angular velocity,
and the problem becomes stationary and axisymmetric. 
Nevertheless, this problem is more complicated than \cite{kfw} 
in that 
both the mass and angular momentum perturbation are involved.

This paper is organized as follows. 
The first step is to obtain the perturbed Weyl scalars $\psi_0$ and $\psi_4$ 
by solving the Teukolsky equation which is discussed in  Section \ref{section:TeuEq}. 
Next in Section \ref{section:CCK}, we describe the CCK formalism in a general form. 
In Section \ref{section:IRG}, the Hertz potential is obtained from $\psi_0$ and $\psi_4$.
In Section \ref{section:PsiP}, we briefly discuss the gravitational fields computed 
from the Hertz potential which contains only $l\geq 2$ modes, and show the presence of the 
singularities in the gravitational fields. 
In Section \ref{section:HertzH}, we obtain the Hertz potential 
of $l=0, 1$ modes by considering the continuity of the gravitational field,
and obtain the metric perturbation from the completed Hertz potential. 
Section \ref{section:summary} is devoted to summary and discussion.

\section{Solutions of the Teukolsky equation}
\label{section:TeuEq}
In this section we analytically derive $\psi_0$ and $\psi_4$. 
The details of the derivation {are}
given in Appendix \ref{section:NP} and \ref{section:solTeukolsky}. 
Here, we only give the outline and the main results which are used in the 
subsequent sections. 

The Schwarzschild metric is given as 
\begin{equation}
\begin{split}
\dev s^2 = -\frac{\Delta}{r^2}\dev t^2+\frac{r^2}{\Delta}\dev r^2+r^2(\dev\theta^2+\sin^2\theta \dev\phi^2)~,
\end{split}
\label{eq:schwametric}
\end{equation}
where $\Delta =r^2-2Mr$~.
Five complex Weyl scalars are defined as
\begin{equation}
\begin{split}
&\Psi_0 = +C_{abcd}l^a m^b l^c m^d~,\\
&\Psi_1 = +C_{abcd}l^a {n}^b l^c {m}^d~,\\
&\Psi_2 = +C_{abcd}l^a m^b \overline{m}^c  n^d~,\\
&\Psi_3 = +C_{abcd}l^a n^b \overline{m}^c n^d~,\\
&\Psi_4 = +C_{abcd}n^a \overline{m}^b n^c \overline{m}^d~,
\label{weylscalar-def}
\end{split}
\end{equation}
{where $C_{abcd}$ is the Weyl tensor, and}
\sout{Here,} 
$l^a,  {n}^b, {m}^d$ are the Kinnersley tetrad defined in Appendix \ref{section:NP}.
The overline $\overline{m}$ denotes the complex conjugate of $m$. 
Note that we adopt the $-$$+$$+$$+$ signature  which is different from that of 
Newman and Penrose \cite{np62} and Teukolsky \cite{teu}. 
Because of it, although 
the sign of above Weyl scalars are opposite from those by Newman and Penrose \cite{np62} 
and Teukolsky \cite{teu}, the Teukolsky equations are left unchanged. 
In the case of Schwarzschild metric, nonzero Weyl scalar is $\Psi_2$. 
\begin{equation}
\Psi_2 
= -\frac{M}{r^3}~.
\end{equation}
The corresponding perturbed Weyl scalars are denoted by $\psi_0,~\psi_1,\ldots,~\psi_4$.

We consider the perturbation of the Schwarzschild metric induced by a rotating ring 
which is composed by a set of point masses in a circular, geodesic orbit
on the equatorial plane. 
The energy-momentum tensor of the ring is written as
\begin{equation}
\begin{split}
T^{ab} =&  \sekibun{}{}{\phi'}\frac{\mass u^a u^b}{u^tr_0{}^2}\delta(r-r_0)\delta(\cos\theta)\delta(\phi-\phi')\\
=& \frac{\mass u^a u^b}{u^tr_0{}^2}\delta(r-r_0)\delta(\cos\theta)~,
\label{Tring}
\end{split}
\end{equation}
where $r_0$ is the radius of the ring, and 
$u^a = u^t\left( (\partial_t)^a+\Omega(\partial_\phi)^a \right)$
{is the four-velocity of the ring}.
The angular velocity $\Omega$ and $u^t$ {are} given as
\begin{equation}
\Omega=\sqrt{\frac{M}{r_0{}^3}}~,~~~~u^t = \sqrt{\frac{r_0}{r_0-3M}}~.
\end{equation}
The rest mass of the ring becomes $2\pi \mass ~(\ll M)$.

Since our {perturbed} space-time is independent from $t$ and $\phi$, 
it is sufficient to consider the case of $\omega=0$ and the $m=0$ mode of 
the spin-weighted spherical harmonics ${}_sY_{lm}(\theta,\phi)$. 
We expand $\psi_0$ as 
\begin{equation}
\begin{split}
&\psi_0(r,\theta) = \sum_{l=2}^{\infty}R^{(2)}_l(r)~{}_{2}Y_{l}(\theta)~.
\label{T2l0}
\end{split}
\end{equation}
The Teukolsky equation for $\psi_0$ is given as
\begin{equation}
\begin{split}
\left[ 
\frac{1}{r^2\Delta^2}\bibun{}{r}{}  \left(\Delta^3\bibun{}{r}{}  \right)-\frac{(l-2)(l+3)}{r^2} 
\right] 
R^{(2)}_l = -4\pi T^{(2)}_l 
~{.}
\label{dokei0}
\end{split}
\end{equation}
We also expand $\psi_4$ as 
\begin{equation}
\begin{split}
& \rho^{-4}\psi_4(r,\theta) = \sum_{l=2}^{\infty}R^{(-2)}_l(r)~{}_{-2}Y_{l}(\theta)~.
\label{T-2l4}
\end{split}
\end{equation}
The Teukolsky equation for $\psi_4$ is given as
\begin{equation}
\begin{split}
\left[ 
\frac{\Delta^2}{r^2}\bibun{}{r}{}  \left(\frac{1}{\Delta}\bibun{}{r}{}  \right)-\frac{(l+2)(l-1)}{r^2} 
\right] 
R^{(-2)}_l = -4\pi T^{(-2)}_l
~{.}
\label{dokei4}
\end{split}
\end{equation}
Here we defined $_sY_l(\theta)$ as
\begin{equation}
_sY_l(\theta)\equiv{}_sY_{l0}(\theta,0)~.
\end{equation}
The source terms $T^{(2)}_l$ and $T^{(-2)}_l$ are given as
\begin{equation}
\begin{split}
T^{(2)}_l
= +2\pi & \frac{1}{r^4} {\mass u^t r_0{}^2}\frac{1}{r^2}\delta(r-r_0)
\\&\times
\sqrt{(l+2)(l-1)(l+1)l} ~{}_{0}Y_l(\pi/2)\\
-2\im \cdot 2\pi & \frac{1}{r^4} {\mass u^t\Omega r_0{}^3} \henbibun{}{r}{}\frac{1}{r} \delta(r-r_0) 
\\&\times 
\sqrt{(l+2)(l-1)}  {}_{1}Y_l(\pi/2)\\
-2\pi & \frac{1}{r^4} {\mass u^t\Omega^2 r_0{}^4}r^2\henbibun{}{r}{}\frac{1}{r^2} \henbibun{}{r}{}  \delta(r-r_0)
\\&\times
  {}_{2}Y_l(\pi/2)~,
\end{split}
\end{equation}
\begin{equation}
\begin{split}
T^{(-2)}_l
= +2\pi & \frac{\Delta^2}{4r^4} {\mass u^t r_0{}^2}\frac{1}{r^2}\delta(r-r_0)
\\&\times
\sqrt{(l+2)(l-1)(l+1)l} ~{}_{0}Y_l(\pi/2)\\
+2\im \cdot 2\pi & \frac{\Delta^2}{4r^4} {\mass u^t\Omega r_0{}^3} \henbibun{}{r}{}\frac{1}{r} \delta(r-r_0) 
\\&\times 
\sqrt{(l+2)(l-1)}  {}_{-1}Y_l(\pi/2)\\
-2\pi & \frac{\Delta^2}{4r^4} {\mass u^t\Omega^2 r_0{}^4}r^2\henbibun{}{r}{}\frac{1}{r^2} \henbibun{}{r}{}  \delta(r-r_0)
\\&\times
  {}_{-2}Y_l(\pi/2)~.
\end{split}
\end{equation}
A simple relation $\frac{\Delta^2}{4}T^{(2)}_l(r)=T^{(-2)}_l(r)$ holds because of the symmetries.

The Teukolsky equations for $\psi_0$ and $\psi_4$ above are solved by {using the} Green's function,
and we obtain
\begin{equation}
\begin{split}
R_l^{(2)} =
&+\frac{4}{M\Delta} \frac{4\pi^2 \mass u^t~{}_{0}Y_l(\pi/2)}{\sqrt{(l+2)(l+1)l(l-1)}}
\\&~~~ \times
\left( -\frac{\Delta_0}{2r_0{}^2}  P_l^2\left(x_0^{<}\right)Q_l^2\left(x_0^{>}\right) \right)
\\%
& -\im \frac{4}{M\Delta} \frac{8\pi^2 \mass u^t \Omega r_0{}^2 ~{}_{-1}Y_l(\pi/2)}{\sqrt{(l+2)(l-1)}(l+1)l}
\\&~~~ \times
\left( -\bibun{}{r_0}{} \frac{\Delta_0}{2r_0{}^2} P_l^2\left(x_0^{<}\right)Q_l^2\left(x_0^{>}\right) \right) 
\\%
& -\frac{4}{M\Delta} \frac{4\pi^2 \mass u^t \Omega^2 r_0{}^4 {}_{-2}Y_l(\pi/2)}{{(l+2)(l-1)(l+1)l}}
\\&~~~ \times
\left( -\bibun{}{r_0}{} \frac{1}{r_0{}^2}
\bibun{}{r_0}{} r_0{}^2 \frac{\Delta_0}{2r_0{}^2}
P_l^2\left(x_0^{<}\right)Q_l^2\left(x_0^{>}\right) \right)  ~,
\end{split}
\label{eq:TeuR2}
\end{equation}
\begin{equation}
\begin{split}
R_l^{(-2)} =
&+\frac{\Delta}{M} \frac{4\pi^2 \mass u^t~{}_{0}Y_l(\pi/2)}{\sqrt{(l+2)(l+1)l(l-1)}}
\\&~~~ \times
\left( -\frac{\Delta_0}{2r_0{}^2}  P_l^2\left(x_0^{<}\right)Q_l^2\left(x_0^{>}\right) \right) 
\\%
& -\im \frac{\Delta}{M} \frac{8\pi^2 \mass u^t \Omega r_0{}^2 ~{}_{-1}Y_l(\pi/2)}{\sqrt{(l+2)(l-1)}(l+1)l}
\\&~~~ \times
\left( -\bibun{}{r_0}{} \frac{\Delta_0}{2r_0{}^2} P_l^2\left(x_0^{<}\right)Q_l^2\left(x_0^{>}\right) \right) 
\\%
& -\frac{\Delta}{M}\frac{4\pi^2 \mass u^t \Omega^2 r_0{}^4 {}_{-2}Y_l(\pi/2)}{{(l+2)(l-1)(l+1)l}}
\\&~~~ \times
\left( -\bibun{}{r_0}{} \frac{1}{r_0{}^2}
\bibun{}{r_0}{} r_0{}^2 \frac{\Delta_0}{2r_0{}^2}
P_l^2\left(x_0^{<}\right)Q_l^2\left(x_0^{>}\right) \right)  ~,
\end{split}
\label{eq:TeuR2m}
\end{equation}
where
\begin{equation}
{\Delta_0 \equiv r_0{}^2 -2Mr_0~,}
\end{equation}
\begin{equation}
x_0^{<}\equiv \frac{{\rm min}(r,r_0)-M}{M}~,~~~~~x_0^{>}\equiv \frac{{\rm max}(r,r_0)-M}{M}~.
\end{equation}
These two radial functions are related as $\frac{\Delta^2}{4}R_l^{(2)}(r) = R_l^{(-2)}(r)$. 
With this relation, together with the fact $_2Y_l(\theta) ={}_{-2}Y_l(\theta)$, 
we find that $\psi_0$ and $\psi_4$ are related in a very simple equation,
\begin{equation}
\psi_4 = \frac{\Delta^2}{4r^4} \psi_0~.
\end{equation}
Note that this relation {holds} because of the symmetries of our space-time. 

\begin{figure*}[htbp]
\begin{center}
\includegraphics[width=8cm]{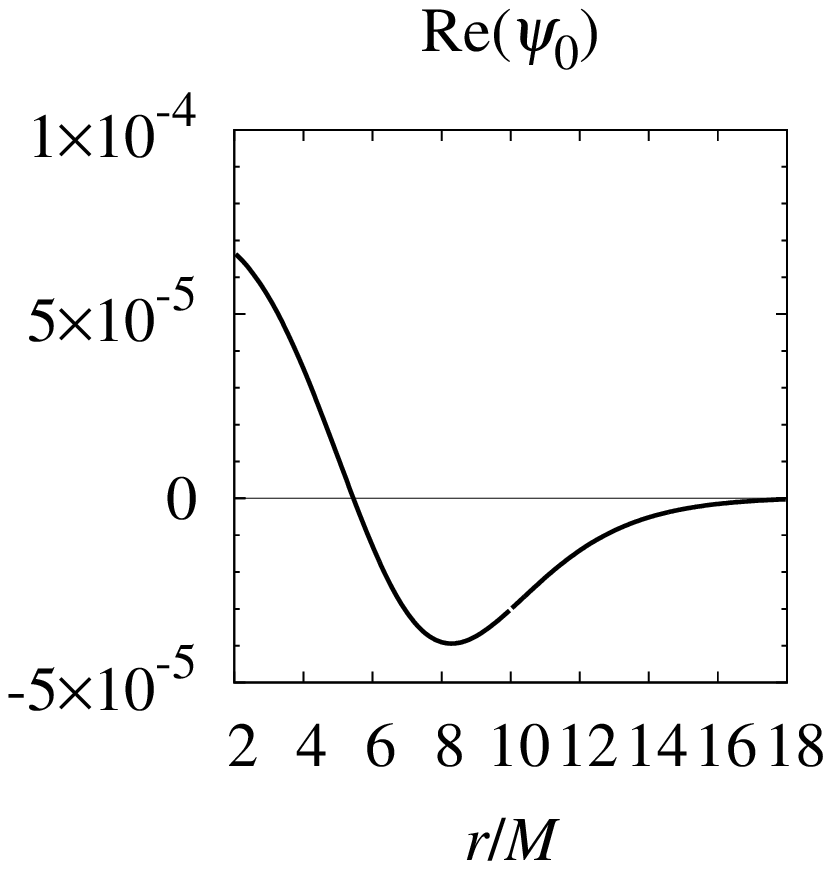}
\includegraphics[width=8cm]{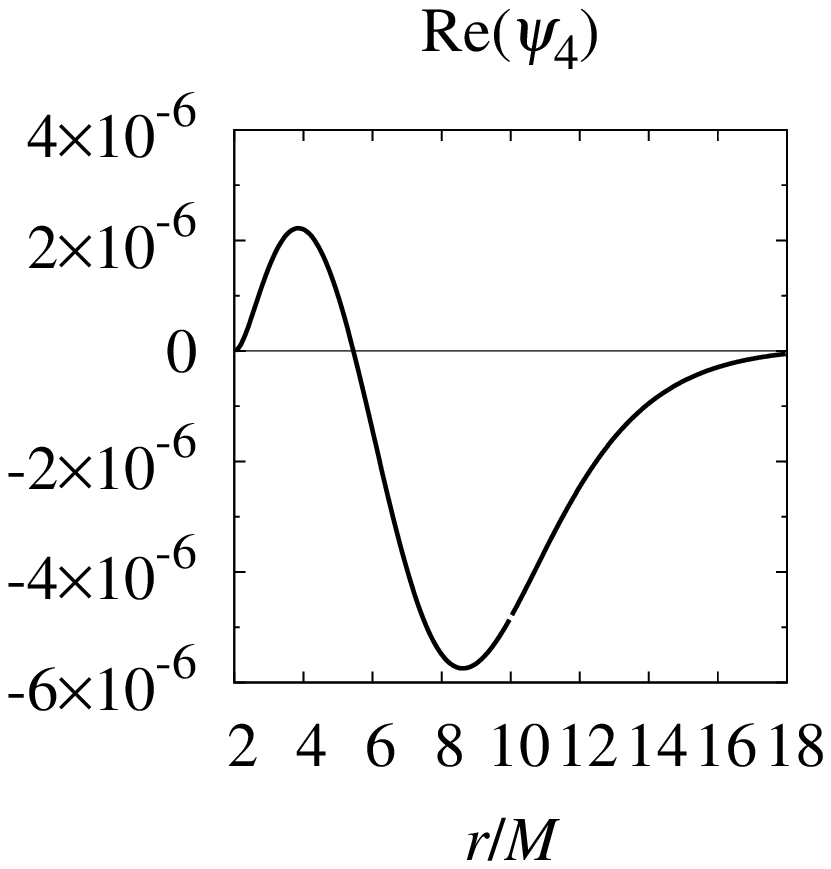}
\\
\includegraphics[width=8cm]{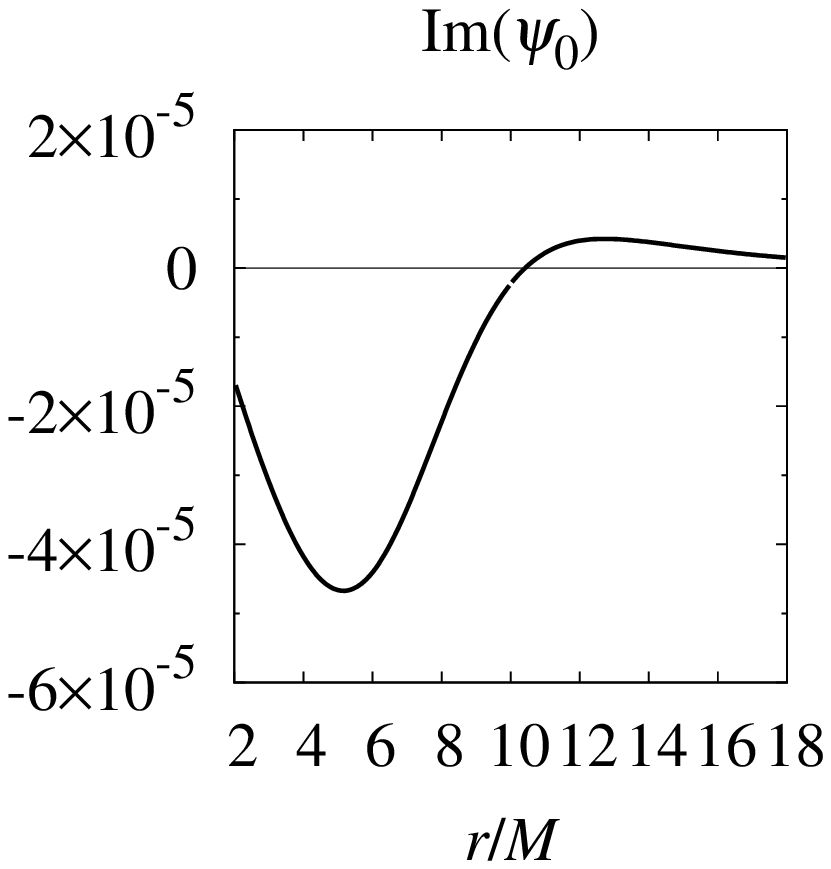}
\includegraphics[width=8cm]{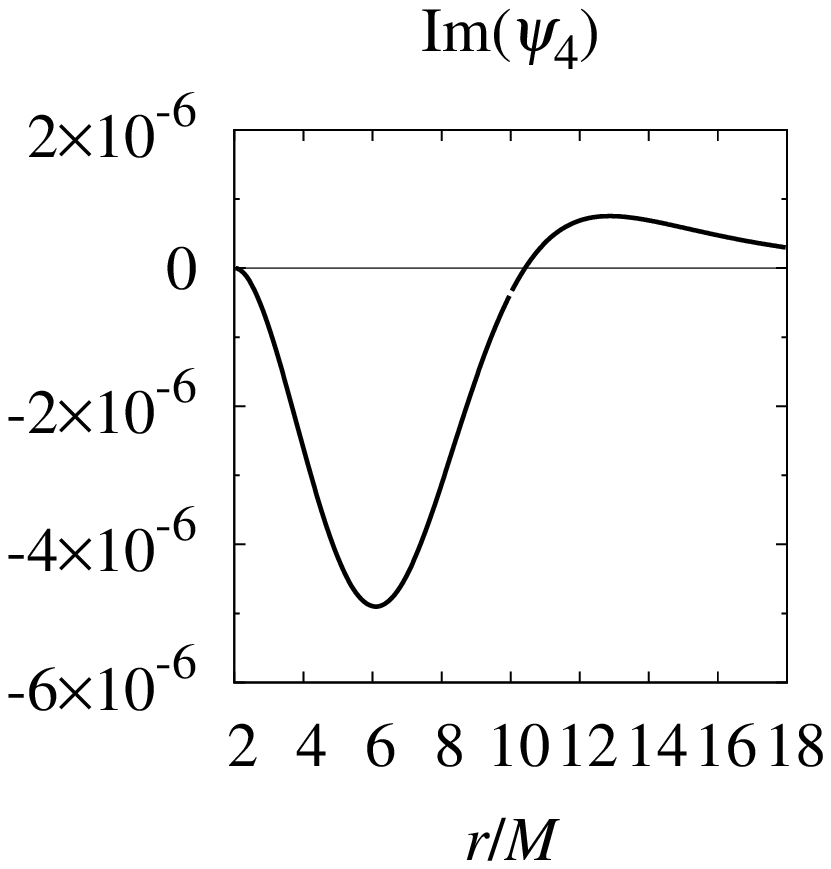}
\end{center}
\caption{Radial dependence of $\psi_0$ and $\psi_4$ obtained by solving the Teukolsky equation. 
The real parts of $\psi_0(r,\theta=\pi/4)$ (top left) and $\psi_4(r,\theta=\pi/4)$ (top right), 
and the imaginary parts of $\psi_0(r,\theta=\pi/4)$ (bottom left) and $\psi_4(r,\theta=\pi/4)$ (bottom right) 
are shown. 
The radius of the ring is $r_0=10M$, and {$m = M/100$}.
We see the smoothness at $r=r_0$.}
\label{fig:TeuSolution}
\end{figure*}

We also find that because the matter is present on the equatorial plane, 
and ${}_sY_l(\theta)$ is evaluated only at $\theta=\pi/2$, we have 
\begin{equation*}
\Re\left(R_l^{(\pm2)}(r)\right)=0~~~{\rm for}~{\rm odd}~l
{~,}
\end{equation*}
and
\begin{equation*}
\Im\left(R_l^{(\pm2)}(r)\right)=0~~~{\rm for}~{\rm even}~l
{~.}
\end{equation*}
Therefore, the real part of $\psi_0$ and $\psi_4$ is symmetric 
about the equatorial plane and the imaginary part is antisymmetric. 
\begin{equation}
\begin{split}
& \Re(\psi_{0/4}(r, \pi-\theta)) = \Re(\psi_{0/4}(r, \theta))~,\\
& \Im(\psi_{0/4}(r, \pi-\theta)) = -\Im(\psi_{0/4}(r, \theta))~.
\label{oddeven}
\end{split}
\end{equation}
In Fig.~\ref{fig:TeuSolution}, 
We show the radial dependence of $\psi_0$ and $\psi_4$, 
with fixed angular coordinate $\theta=\pi/4$. 
Note that $\psi_0$ and $\psi_4$ are smooth at the sphere, $r=r_0$,
except for $\theta=\pi/2$, where the energy-momentum tensor vanishes.
%


%

\section{Construction of the perturbed gravitational fields}
\label{section:Construction}
{Chrzanowski \cite{c} and Cohen and Kegeles \cite{kc} introduced a formalism }
to compute the perturbed
metric in a ``radiation gauge'' from Teukolsky valuables $\psi_0$ and $\psi_4$. 
{In this section, we describe how we can use the CCK formalism to calculate the perturbed gravitational 
fields produced by the rotating ring. }

\subsection{The CCK formalism}
\label{section:CCK}
{In the CCK formalism, the Hertz potential $\Psi$,  which is a solution of the homogeneous Teukolsky equation, 
is introduced. The perturbed metric is obtained by differentiating the Hertz potential. 
In order to obtain the relation between the Hertz potential and the perturbed metric, 
two kinds of gauge conditions are used. 
They are called ``Ingoing Radiation Gauge'' (IRG) and ``Outgoing Radiation Gauge'' (ORG).}
The IRG is defined by the conditions $h_{ab}l^b=h^a{}_a=0$. 
The perturbed metric $h_{ab}$ in IRG is related to the Hertz potential as
\begin{equation}
\begin{split}
h_{ab} &= -\big[ l_a l_b(\overline{\pmb\delta} +2\beta)(\overline{\pmb\delta} +4\beta)\overline{\Psi}\\
&~~~~~ -2l_{(a}\overline{m}_{b)}(\pmb D +\rho)(\overline{\pmb\delta} +4\beta)\overline{\Psi} \\
&~~~~~~~~~~ +\overline{m}_a\overline{m}_b(\pmb D -\rho) (\pmb D +3\rho)\overline{\Psi} \big] \\
&~~~ + \left[ {\rm c.c.} \right]~,
\label{irg}
\end{split}
\end{equation}
%
where $\left[ {\rm c.c.} \right]$ represents the complex conjugate of the first term. 
The bold greek characters are derivative operators associated 
with the tetrad defined in Appendix \ref{section:NP}. 
The Hertz potential $\Psi$ in IRG satisfies the source-free Teukolsky equation 
{with $s=-2$. }
\begin{equation}
(\pmb{\Delta} +\mu +2\gamma)({\pmb D} +3\rho)\Psi -3\Psi_2\Psi 
= (\overline{\pmb \delta} -2\beta) ({\pmb \delta} +4\beta) \Psi~.
\label{IRGhomo1}
\end{equation}
Equivalently, this equation is written as
\begin{equation}
(\pmb{\Delta} -2\mu +2\gamma)\pmb{D} \Psi +3\rho\partial_t \Psi 
= (\overline{\pmb{\delta}} -2\beta)(\pmb{\delta} +4\beta)\Psi~.
\label{IRGhomo}
\end{equation}

By using \eref{irg} and \eref{IRGhomo1}, 
the relations between the perturbed Weyl scalars and the Hertz potential are obtained as
\cite{ref:KFWtypo}
\begin{subequations}
\label{weyl-hertz}
\begin{align}
\psi_0 &= \frac{1}{2} \pmb D^4\overline{\Psi}~, \label{eq:psi0Hertz1} \\
\psi_1 &= \frac{1}{2} \pmb D^3(\overline{\pmb\delta} +4\beta)\overline{\Psi}~,\\
\psi_2 &= \frac{1}{2} \pmb D^2(\overline{\pmb\delta} +2\beta)(\overline{\pmb\delta} +4\beta)\overline{\Psi}~,\\
\psi_3 &= \frac{1}{2} \pmb D \overline{\pmb\delta}(\overline{\pmb\delta} +2\beta)(\overline{\pmb\delta} +4\beta)\overline{\Psi} + 3\gamma\pmb{D}\rho(\pmb{\delta} +4\beta)\Psi~, \label{eq:psi3Hertz}
\\
\psi_4 &= \frac{1}{2}(\overline{\pmb\delta} -2\beta)\overline{\pmb\delta}(\overline{\pmb\delta} +2\beta)(\overline{\pmb\delta} +4\beta)\overline{\Psi} -3\gamma\rho^2 \partial_t\Psi~. \label{eq:psi4Hertz1}
\end{align}
\end{subequations}
%

On the other hand, ORG is defined by the conditions $h_{ab}n^b=h^a{}_a=0$. 
The perturbed metric $h_{ab}^{\rm ORG}$ {is related to the Hertz potential as}
\begin{equation}
\begin{split}
h_{ab}^{\rm ORG} &= -\Big[ n_a n_b\left(-\frac{2r^2}{\Delta}\right)^2 (\overline{\pmb\delta} +2\beta)(\overline{\pmb\delta} +4\beta)\frac{\Delta^2}{4} {\Psi}\\
&~~~~~ -2 n_{(a}\overline{m}_{b)}\left(-\frac{2r^2}{\Delta}\right)(\pmb D +\rho)(\overline{\pmb\delta} +4\beta)\frac{\Delta^2}{4} {\Psi} \\
&~~~~~~~~~~ +\overline{m}_a\overline{m}_b(\pmb D -\rho) (\pmb D +3\rho)\frac{\Delta^2}{4} {\Psi} \Big] \\
&~~~+\left[ {\rm c.c.} \right]~.
\label{org}
\end{split}
\end{equation}
The Hertz potential $\Psi$ in ORG satisfies the source-free Teukolsky equation {with $s=2$}. 
\begin{equation}
\begin{split}
(\tilde{\pmb\Delta} +\mu +2\gamma)&(\tilde{\pmb{D}} +3\rho)\frac{\Delta^2}{4}\Psi
-3\Psi_2\frac{\Delta^2}{4}\Psi 
\\&~~~~~
= (\pmb{\delta} -2\beta)(\overline{\pmb{\delta}} +4\beta)\frac{\Delta^2}{4}\Psi~.
\label{HertzhomoORG}
\end{split}
\end{equation}
Equivalently, this equation is written as
\begin{equation}
\begin{split}
(\tilde{\pmb\Delta} -2\mu +2\gamma)\tilde{\pmb D} \frac{\Delta^2}{4}\Psi &
- 3\rho\partial_t \frac{\Delta^2}{4}\Psi \\
&= (\pmb\delta -2\beta)(\overline{\pmb\delta} +4\beta)\frac{\Delta^2}{4}\Psi~.
\end{split}
\end{equation}

By using \eref{org} and \eref{HertzhomoORG}, 
the relations between the perturbed Weyl scalars and the Hertz potential are {obtained} as
\begin{subequations}
\label{weyl-hertz-org}
\begin{align}
\left(-\frac{2r^2}{\Delta}\right)^2
\psi_4 
&= \frac{1}{2} \pmb D^4\frac{\Delta^2}{4}\overline{\Psi}~, \label{eq:psi0Heltz2} \\
\left(-\frac{2r^2}{\Delta}\right)
\psi_3 
&= \frac{1}{2} \pmb D^3(\overline{\pmb\delta} +4\beta)\frac{\Delta^2}{4}\overline{\Psi}~,\\
\psi_2 
&= \frac{1}{2} \pmb D^2(\overline{\pmb\delta} +2\beta)(\overline{\pmb\delta} +4\beta)\frac{\Delta^2}{4}\overline{\Psi}~,\\
\left(-\frac{\Delta}{2r^2}\right)
\psi_1 
&= \frac{1}{2} \pmb D \overline{\pmb\delta}(\overline{\pmb\delta} +2\beta)(\overline{\pmb\delta} +4\beta)\frac{\Delta^2}{4}\overline{\Psi} 
\nonumber\\&~~~~~~~~~~
+ 3\gamma\pmb{D}\rho(\pmb{\delta} +4\beta) \frac{\Delta^2}{4}\Psi~,
\\
\left(-\frac{\Delta}{2r^2}\right)^2 
\psi_0 
&= \frac{1}{2}({\pmb\delta} -2\beta){\pmb\delta}({\pmb\delta} +2\beta)({\pmb\delta} +4\beta)\frac{\Delta^2}{4}\overline{\Psi} 
\nonumber\\&~~~~~~~~~~
+3\gamma\rho^2 \partial_t \frac{\Delta^2}{4}\Psi~. \label{eq:psi4Heltz2}
\end{align}
\end{subequations}

Whichever gauge we choose, we look for the Hertz potential that satisfies 
{the relations to $\psi_0$ and $\psi_4$,
Eqs. \eref{eq:psi0Hertz1} and \eref{eq:psi4Hertz1}, 
or Eqs. \eref{eq:psi0Heltz2} and \eref{eq:psi4Heltz2}.}

\subsection{The Hertz potential and the metric perturbation in IRG}
\label{section:IRG}
In this paper, we use IRG to construct the perturbed gravitational fields. 
From \eref{weyl-hertz}, the relations between Teukolsky valuables and the Hertz potential become
\begin{eqnarray}
&&\psi_0 =
\frac{1}{2}\left(\frac{\partial}{\partial r}\right)^4\overline{\Psi}~,
\label{psi0-hertz}
\\
&&\psi_4  =
\frac{1}{2}\frac{1}{4r^4}\sin^2\theta \left(\frac{\partial}{\partial \cos\theta}\right)^4 \sin^2\theta \overline{\Psi} ~.
\label{psi4-hertz}
\end{eqnarray}
{Here,} we used the fact that the ring and the black hole are {stationary} and axisymmetric.

Our task is to find Hertz potential which satisfies \eref{psi0-hertz}, \eref{psi4-hertz} and \eref{IRGhomo}. 

{By} substituting the solution of the Teukolsky {equation,}
\begin{equation*}
\psi_4 = \frac{1}{r^4} \sum_{l=2}^\infty R_l^{(-2)}(r)_{-2}Y_l(\theta)
\end{equation*}
into \eref{psi4-hertz}, {we obtain}
\begin{equation}
\sum_{l=2}^{\infty} 8 R_l^{(-2)} \frac{_{-2}Y_l(\theta)}{\sin^2\theta} 
= \left(\frac{\partial}{\partial \cos\theta}\right)^4 \sin^2\theta \overline{\Psi}~{.}
\end{equation}
From \eref{spinup}, we can obtain the following relation 
\begin{equation}
\left(\frac{\partial }{\partial \cos\theta}\right)^4 \frac{_{-2}Y_l(\theta)}{\sin^2\theta} 
= \frac{1}{\sin^2\theta} \frac{_2Y_l(\theta)}{(l+2)(l-1)(l+1)l}
~{.}
\end{equation}
By using this relation, $\Psi$ can be integrated as
\begin{equation}
\begin{split}
\overline{\Psi}(r,\theta) &= \overline{\Psi_{\rm P}}+\overline{\Psi_{\rm H}}, 
\label{hertzPH}
\end{split}
\end{equation}
where
\begin{equation}
\overline{\Psi_{\rm P}} \equiv \sum_{l=2}^{\infty} \frac{8{R_l^{(-2)}(r) _{2}Y_l(\theta)}}{(l+2)(l-1)(l+1)l} ~,
\label{PsiP}
\end{equation}
\begin{equation}
\begin{split}
\overline{\Psi_{\rm H}} &\equiv \frac{2A}{\sin^2\theta} \bigg( \frac{a(r)}{6}\cos^3\theta +\frac{b(r)}{2}\cos^2\theta
\\&~~~~~~~~~~~~~~~~~~~~~~~~~
 + c(r)\cos\theta +d(r) \bigg)~,
\end{split}
\label{PsiHabcd}
\end{equation}
and where $\overline{\Psi}$ is the complex conjugate of $\Psi$.
$a(r)$, $b(r)$, $c(r)$, and $d(r)$ are arbitrary functions and $A$ is a constant defined as
\begin{equation}
A\equiv \frac{\mass}{r_0\sqrt{\Delta_0}}~.
\end{equation}
Here, $\Psi_{\rm P}$ and $\Psi_{\rm H}$ are the particular solution and the homogeneous solution 
of the equation \eref{psi4-hertz}{,} respectively. 
The particular solution $\Psi_{\rm P}$ satisfies \eref{psi0-hertz} and \eref{IRGhomo} in the region, 
$r\neq r_0$. The reason is as follows.
From the Teukolsky--Starobinsky relation, we obtain
\begin{equation}
\left(\frac{\partial}{\partial r}\right)^4
\frac{R_l^{(-2)}(r)}{(l+2)(l-1)(l+1)l}
= \frac{1}{4}R_l^{(2)}(r)
~{.}
\end{equation}
By using this, we can obtain $\psi_0$ by substituting $\Psi_{\rm P}$ into \eref{psi0-hertz}. 
Further, since $R_l^{(-2)}(r)$ is the solution of the radial Teukolsky equation with the source term
consisting of a circular rotating ring, it satisfies the homogeneous Teukolsky equation
in the region, $r\neq r_0$. 
Thus, it is clear that the particular solution $\Psi_{\rm P}$ of the form \eref{PsiP} satisfies 
\eref{IRGhomo}  in the region, $r\neq r_0$. 
It is now shown that $\Psi_{\rm P}$ is a Hertz potential that satisfies \eref{psi0-hertz}, \eref{psi4-hertz}, and \eref{IRGhomo} 
everywhere except for the region, $r=r_0$. 

$\Psi_{\rm P}$ is not only singular at $r=r_0$, but also does not include lower modes
($l=0, 1$). 
The monopole perturbation and the dipole perturbation of the space{-}time are considered 
to be included in the ``homogeneous solution'' part $\Psi_{\rm H}$.

We can obtain constraints on the functions $a(r)$, $b(r)$, $c(r)$, and $d(r)$ in $\Psi_{\rm H}$ 
from \eref{IRGhomo}.
By substituting $\Psi_{\rm H}$ into \eref{IRGhomo}, we obtain
\begin{equation}
(\pmb{\Delta} -2\mu +2\gamma)\pmb{D} \Psi_{\rm H} 
= (\overline{\pmb{\delta}} -2\beta)(\pmb{\delta} +4\beta)\Psi_{\rm H}~.
\end{equation}
This condition implies that each of $a(r)$, $b(r)$, $c(r)$, and $d(r)$ must be in the following forms. 
\begin{equation}
\begin{split}
& a(r)=a_1 r^2 (r-3M)+a_2~,\\
& b(r)=b_1 r^2+b_2(r-M)~,\\
& c(r)=-\frac{a_1}{2}(r^2+4M^2)(r-M)-\frac{a_2}{2}
\\&~~~~~~~~~~~~~~~ 
+c_1 r^2+c_2(r-M)~,\\
& d(r)=\frac{b_1}{2}r^2 +\frac{b_2}{2}r +d_1 r^2(r-3M)+d_2~.
\label{abcd}
\end{split}
\end{equation}
Here $a_1$, $a_2$, etc. are arbitrary complex constants. 
Then, the right-hand side of \eref{psi0-hertz} vanishes when we substitute $\Psi_{\rm H}$ with constrains \eref{abcd}.
Thus, $\Psi_{\rm H}$ with \eref{abcd} is a homogeneous solution of \eref{psi0-hertz} and \eref{psi4-hertz}, 
and satisfies \eref{IRGhomo}.

It is known \cite{ori03} that
the Hertz potential that globally satisfies \eref{psi0-hertz}, 
\eref{psi4-hertz}, and \eref{IRGhomo} simultaneously does not exist because of the presence of matter (the ring). 
{Thus, we need to give up the global regularity of the solution.
We find that we can obtain a solution which is smooth at $r=r_0$ if we abandon the smoothness of 
the Hertz potential at $(r\geq r_0, ~\theta=\pi/2)$. 
We also find that in order to obtain the smoothness at $r=r_0$, 
we need to include the contribution from the lower modes ($l=0, 1$). 
We show that this can be done by choosing eight complex parameters, $a_1$, $a_2$, etc., 
appropriately, and making the Hertz potential $\Psi=\Psi_{\rm P}+\Psi_{\rm H}$ satisfy \eref{psi0-hertz}, 
\eref{psi4-hertz}, and \eref{IRGhomo} everywhere except for the region $(r\geq r_0,~\theta=\pi/2)$.}

\subsection{Fields corresponding to $\Psi_{\rm P}$}
\label{section:PsiP}
\begin{figure*}[ht]
\begin{center}
\includegraphics[width=5.6cm]{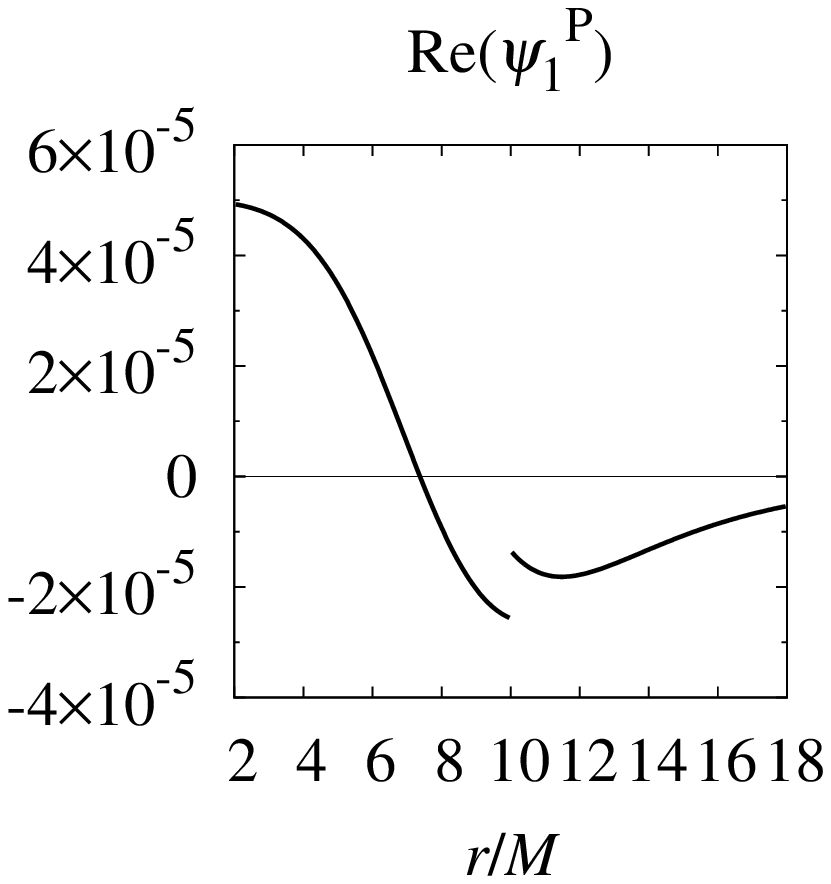}
\includegraphics[width=5.6cm]{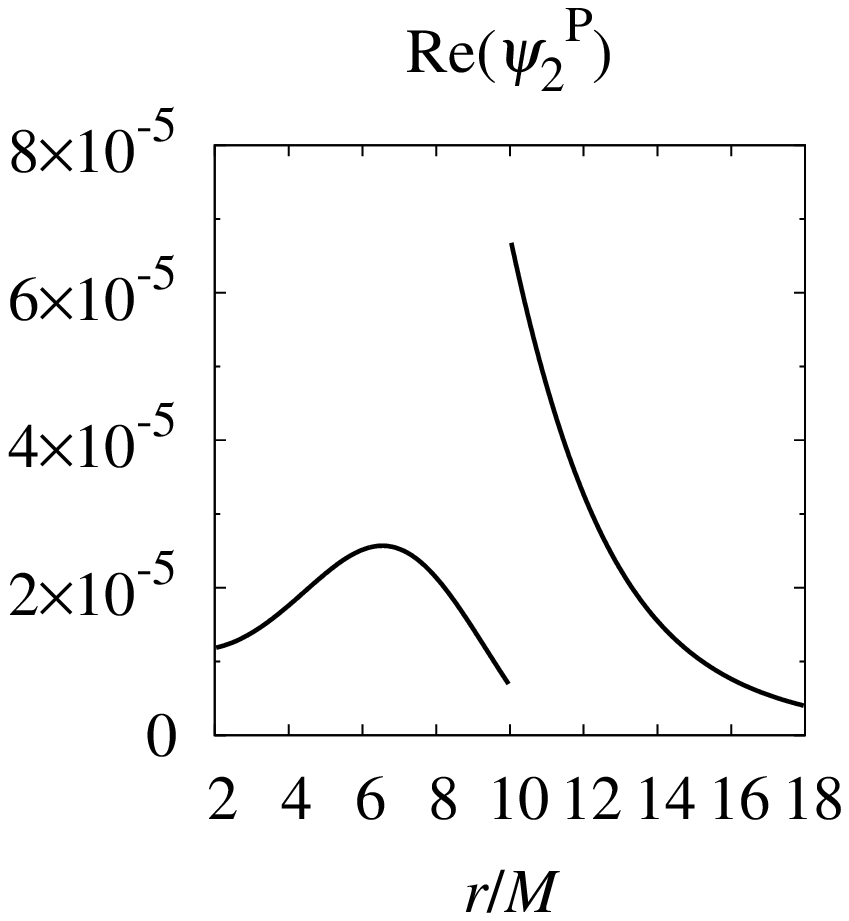}
\includegraphics[width=5.6cm]{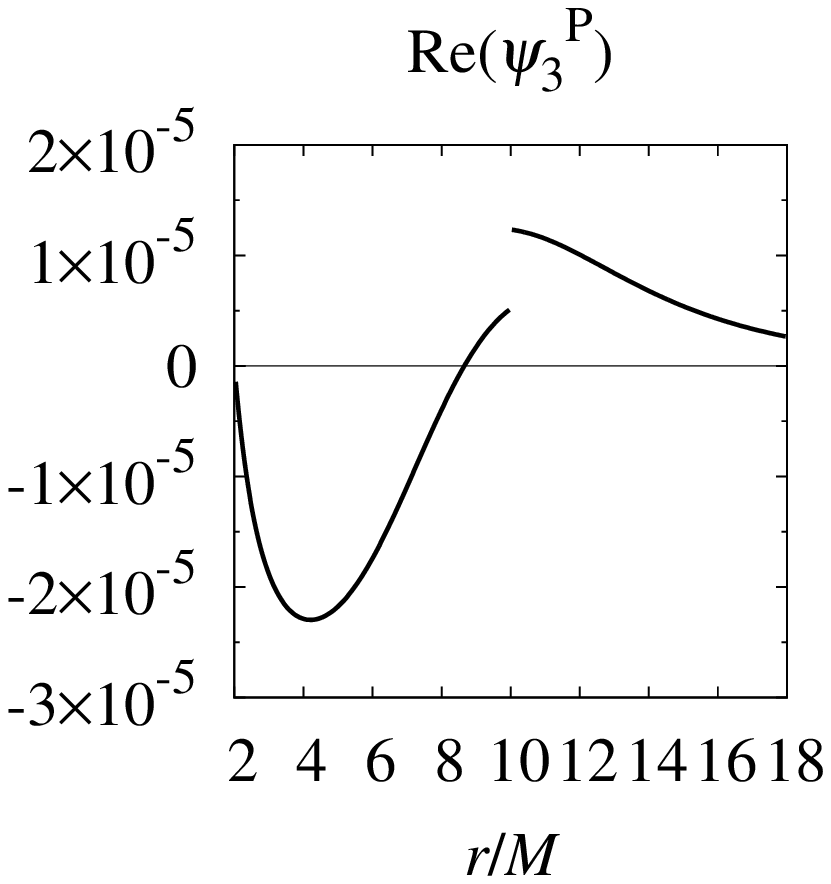}
\end{center}
\caption{Radial dependenc{e} of the real parts of $\psi_1$ (left), $\psi_2$ (center), and $\psi_3$ (right) {derived from}  $\Psi_{\rm P}$ at $\theta=\pi/4$. The radius of the ring is $r_0=10M$. They are discontinuous at $(r=r_0, ~\theta=\pi/4)$.}
\label{fig:Repsi123P}
\end{figure*}
\begin{figure*}[htb]
\begin{center}
\includegraphics[width=5.6cm]{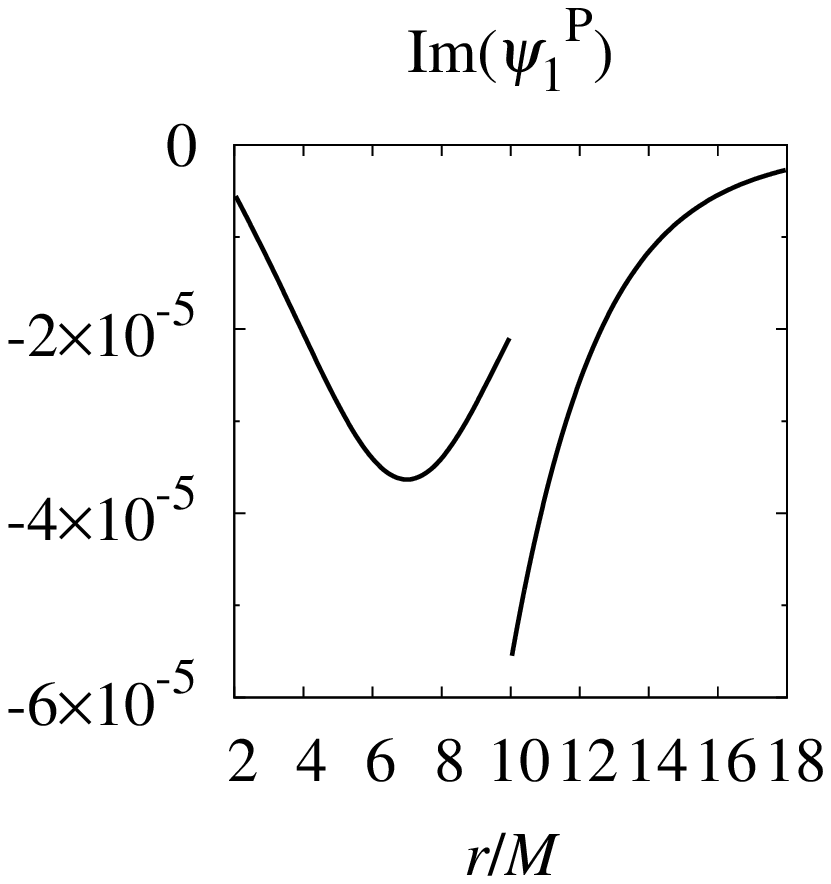}
\includegraphics[width=5.6cm]{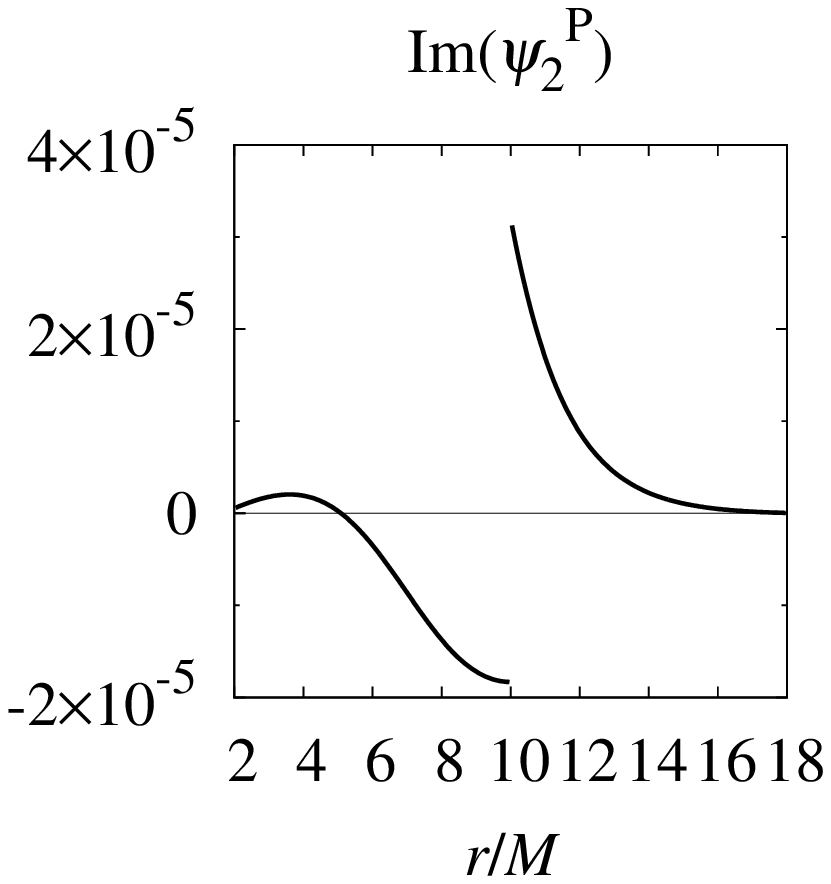}
\includegraphics[width=5.6cm]{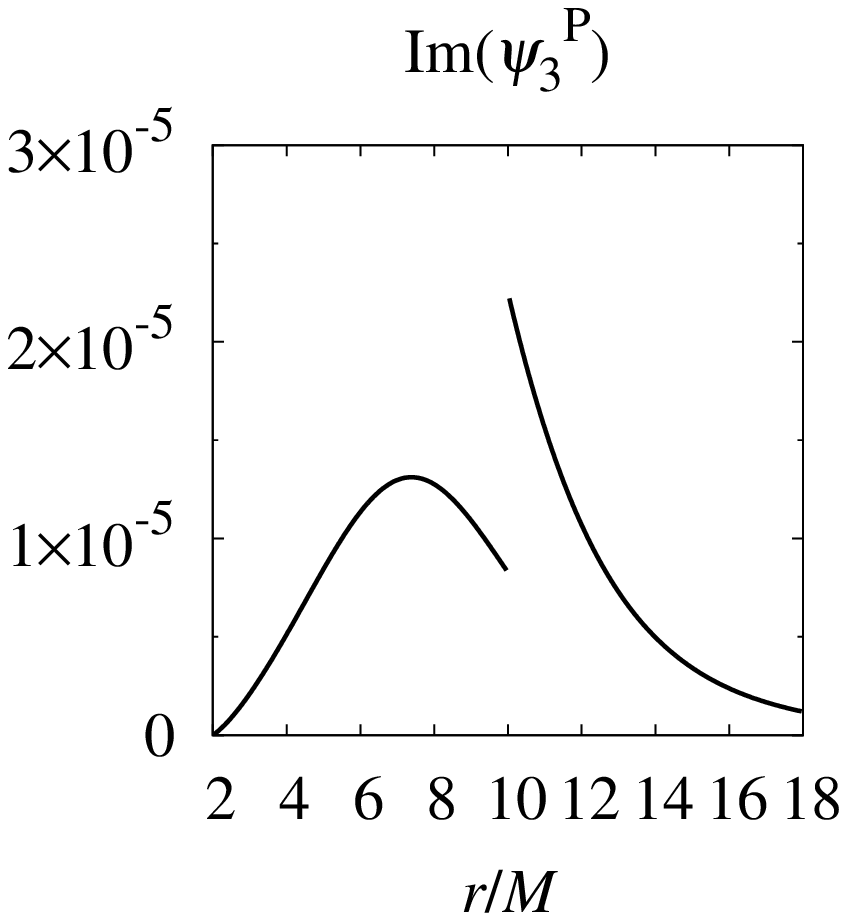}
\end{center}
\caption{Radial dependenc{e} of the imaginary parts of $\psi_1$ (left), $\psi_2$ (center), and $\psi_3$ (right) {derived from}  $\Psi_{\rm P}$ at $\theta=\pi/4$. The radius of the ring is $r_0=10M$. They are discontinuous at $(r=r_0, ~\theta=\pi/4)$.}
\label{fig:Impsi123P}
\end{figure*}
{Here, we demonstrate the behavior of the Weyl scalars associated with $\Psi_{\rm P}$.
We introduce a notation like $\psi_1^{\rm P}$ which means that 
it is calculated by substituting $\Psi=\Psi_{\rm P}$ into the equation for $\psi_1$ in \eref{weyl-hertz}. 
In Figs. \ref{fig:Repsi123P} and \ref{fig:Impsi123P}, we show the radial dependence of the real and 
imaginary parts of $\psi_1^{\rm P}, \psi_2^{\rm P}$ and $\psi_3^{\rm P}$ at $\theta=\pi/4$. }

{As discussed in the previous section, }
$\psi_0^{\rm P}$ agree with the Teukolsky solution $\psi_0$, 
therefore the graph is the same as Fig. \ref{fig:TeuSolution}.
{Other Weyl scalars, }
$\psi_1^{\rm P}$, $\psi_2^{\rm P}$, and $\psi_3^{\rm P}$, have discontinuity on the 
surface of sphere at radius $r=r_0$,  although there is no matter field 
on the surface $(r_0,~\theta\neq \pi/2)$. 
{It is also apparent that the }
perturbed metric $h_{\mu\nu}^{\rm P}$ {calculated from $\Psi_{\rm P}$} 
is not smooth on the surface of 
the sphere, too. 
\begin{figure*}[ht]
\begin{center}
\includegraphics[width=5.6cm]{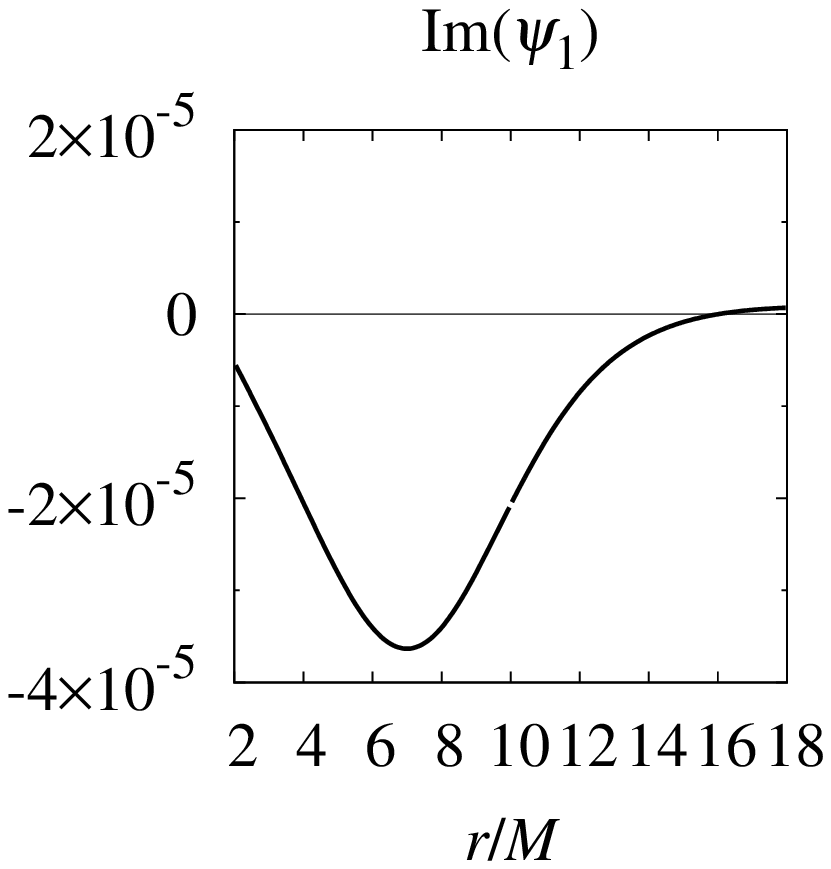}
\includegraphics[width=5.6cm]{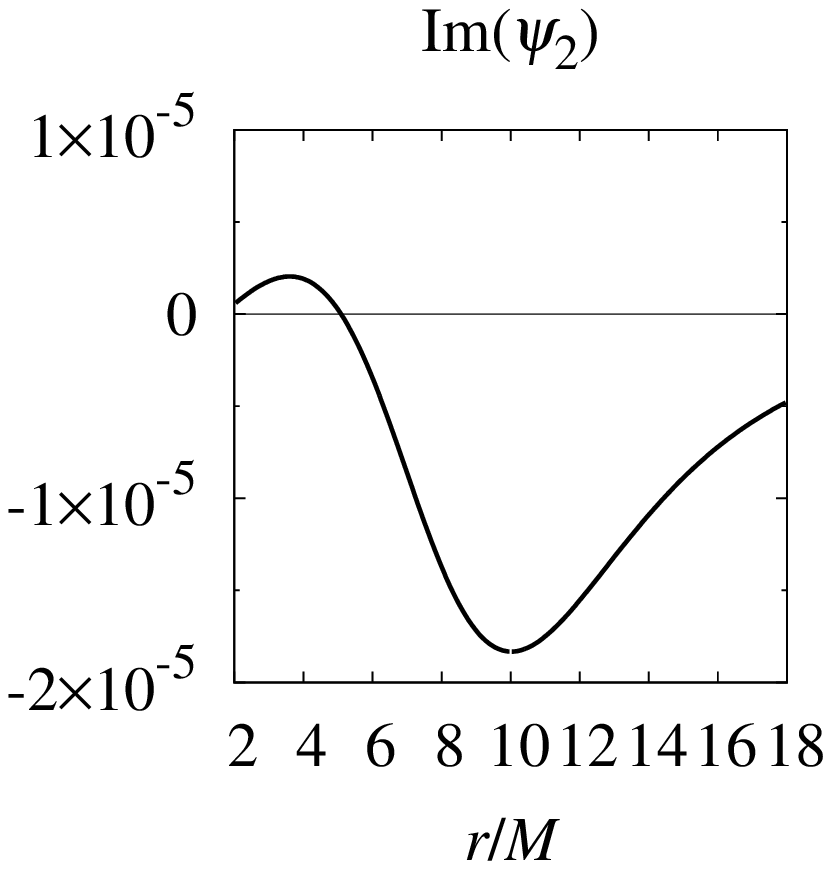}
\includegraphics[width=5.6cm]{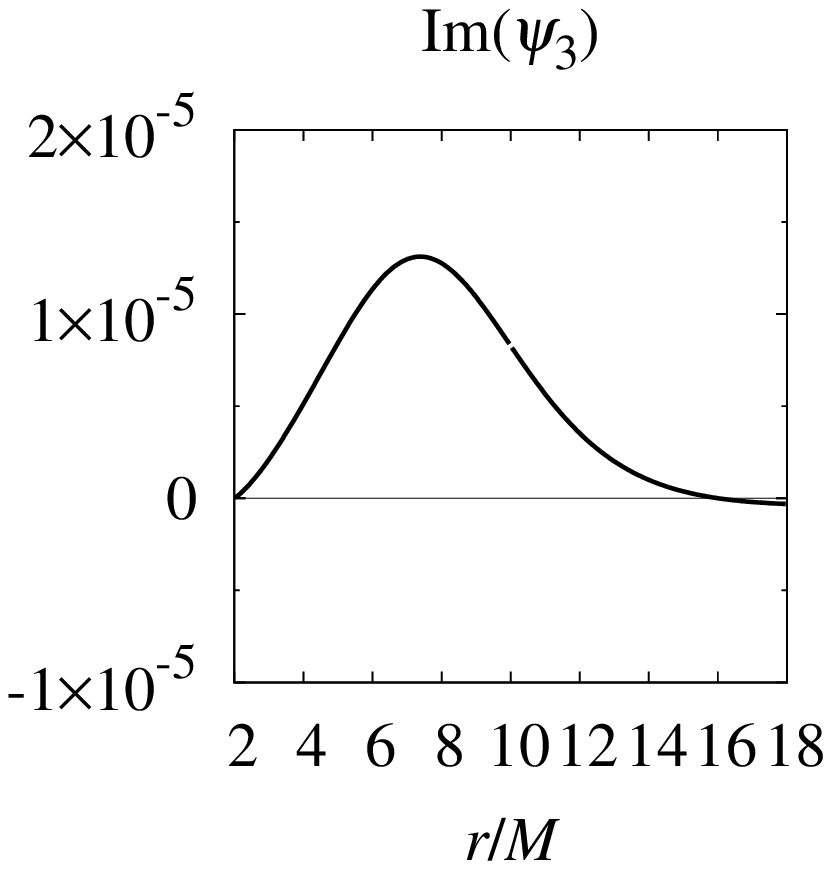}
\end{center}
\caption{Radial dependence of the imaginary parts of $\psi_1$ (left), $\psi_2$ (center), and $\psi_3$ (right) {derived from} $\Psi_{\rm P}+\Psi_{\rm H}$ at $\theta=\pi/4$. The radius of the ring is $r_0=10M$. {It is clear that they are continuous at $r=r_0$.}}
\label{fig:Impsi123}
\end{figure*}

\subsection{$\Psi_{\rm H}$}
\label{section:HertzH}
\subsubsection{{Contribution of angular momentum perturbation}}
Keidl, Friedman, and Wiseman (2007) \cite{kfw} illuminated 
that some of parameters are physical parameters and others are pure gauge. 
They found that $\Re(b_1)$ and $\Re(b_2)$ contribute to {the} mass perturbation of 
the space{-}time and $\Im(a_2)$ contributes to {the} angular momentum perturbation of 
the space{-}time. Specifically, it is found that
\begin{equation}
\begin{split}
& \delta M = -A(3M\Re(b_1)+\Re(b_2))~,
\\
& \delta J = -A\Im(a_2)~.
\label{deltaMdeltaJ}
\end{split}
\end{equation}
The latter relation is obtained as below \cite{kfw}. 
The metric perturbation due to small 
angular momentum to the Schwarzschild space-time is given in the Boyer{--}Lindquist coordinates as
\begin{equation}
h_{ab}^{\rm Kerr} = -\frac{4\delta J}{r}\sin^2\theta (\dev t)_{(a}(\dev\phi)_{b)}
~{.}
\end{equation}
The corresponding tetrad components are
\begin{equation}
h_{23}^{\rm Kerr} = -\im \frac{ \delta J}{\sqrt{2} r^2}\sin\theta~,~~~~~
h_{13}^{\rm Kerr} = -\im \frac{2\delta J}{\sqrt{2}\Delta}\sin\theta~.
\end{equation}
We can transform these into ingoing radiation gauge, with the gauge vector
\begin{equation*}
\xi^a = \xi^3 m^a +\xi^4\overline{m}^a;
\end{equation*}
\begin{equation}
\xi^3 = -\xi^4 = -\frac{\im \delta J}{\sqrt{2}M}
\left( 1+\frac{r}{2M}\ln\left(1-\frac{2M}{r}\right) \right)~.
\end{equation}
The resultant nonzero component of $h_{ab}=h_{ab}^{\rm Kerr}+\mathcal{L}_\xi g_{ab}$ is 
\begin{equation}
h_{23} = -\im \frac{\sqrt{2}\delta J}{r^2}\sin\theta~.
\end{equation}

{The metric associated with the imaginary part of $a_2$ can be obtained 
by inserting \eref{PsiHabcd} and \eref{abcd} into \eref{irg}, and becomes
$h_{23}^{\rm H}=\im (\sqrt{2}A\Im(a_2)/r^2)\sin\theta$. 
We thus obtain $\delta J = -A\Im(a_2)$. }


In our case, $\delta M$ and $\delta J$ are the energy and angular momentum 
of the rotating ring, respectively. They are
\begin{equation}
M_{\rm ring} \equiv -2\pi\mass u_a (\partial_t)^a~,~~~~~J_{\rm ring} \equiv  2\pi\mass u_a (\partial_\phi)^a~, 
\label{MJring}
\end{equation}
where $u^a$ is the four-velocity of the ring,
\begin{equation*}
u^a = \sqrt{\frac{r_0}{r_0-3M}}\left( (\partial_t)^a +\sqrt{\frac{M}{r_0{}^3}}(\partial_\phi)^a \right)~.
\end{equation*}

Interestingly, the jumps of $\Im(\psi_1)$, $\Im(\psi_2)$, and $\Im(\psi_3)$ disappeared
when we choose $\Im(a_2)=0$ for $r<r_0$ and $\Im(a_2)=-\delta J/A$ for $r>r_0$. 
Namely, the imaginary parts of $\psi_1$, $\psi_2$, and $\psi_3$ {are continuous at $r=r_0$}
if we choose 
\begin{eqnarray}
\Psi = \left\{ \begin{array}{ll}
    \Psi_{\rm P}, & (2M < r < r_0)~ \\
    \Psi_{\rm P} + \frac{2\im\delta J}{\sin^2\theta}\left( \frac{1}{6}\cos^3\theta -\frac{1}{2}\cos\theta \right). &(r_0 < r)~
  \end{array} \right. 
\nonumber\\
\label{eq:Psiangmom}
\end{eqnarray}
Further, they also look smooth at $r=r_0$ (Fig. \ref{fig:Impsi123}).

{Although we want to determine other parameters in a similar way,
we can not do it. 
One reason is that since the mass perturbation in \eref{deltaMdeltaJ} contains two parameters, 
$\Re(b_1)$ and $\Re(b_2)$, it is not possible to determine them from only one equation. 
Further, we don't have similar equations for other parameters 
which are not related to the mass and angular momentum perturbation. }

\subsubsection{Determination of all parameters in $\Psi_{\rm H}$}
We now determine all other parameters so that the discontinuity of all the fields at $r=r_0$
{disappears}.

\begin{figure}[ht]
  \begin{center}
\includegraphics[width=8cm]{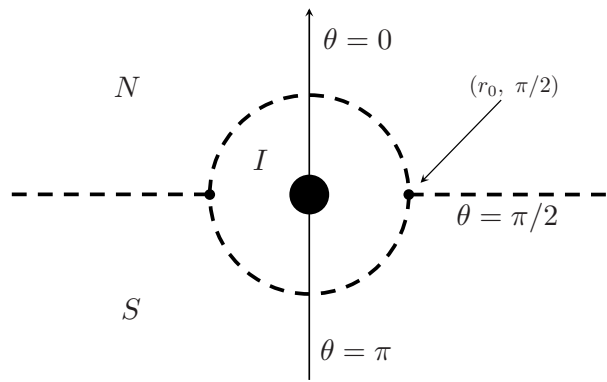}
    \caption{$r$-$\theta$ plane. The three regions are divided by dashed lines. 
    The filled black circle at the center is {the region within the} event horizon of the black hole. 
 The two black dots represent the position of the ring. }
    \label{fig:regions}
  \end{center}
\end{figure}

Details are in the appendix. 
First, we obtain four conditions by demanding that the metric perturbation and the Weyl scalars 
should not diverge at $\theta=0$ and ${\theta=}\pi$. This can be satisfied when the Hertz potential $\Psi$
does not diverge at $\theta=0$ and $\theta=\pi$. 
From the condition at $\theta=0$, we obtain
\begin{equation}
\begin{split}
& 3 d_1 = a_1~,~~~c_1=Ma_1-b_1~,\\
& c_2=2M^2a_1-b_2~,~~~6d_2=2a_2-3Mb_2
~{.}
\label{fromNpole}
\end{split}
\end{equation}
From the condition at $\theta=\pi$, we obtain
\begin{equation}
\begin{split}
& 3 d_1 = -a_1~,~~~c_1=Ma_1+b_1~,\\
& c_2=2M^2a_1+b_2~,~~~6d_2=-2a_2-3Mb_2
~{.}
\label{fromSpole}
\end{split}
\end{equation}
These sets of conditions are simultaneously satisfied if and only if 
$a_1=a_2=b_1=b_2=c_1=c_2=d_1=d_2=0$, i.e. $\Psi_{\rm H}=0$. 
This means that we can not have the contribution from the mass and the angular momentum 
perturbation. 
This implies that we can not obtain the regular solution globally. 
However, we find that if we divide the space{-}time into several region,
we can obtain regular solution in each region. 
{Namely, we divide the region into three regions: 
$(2M<r<r_0)$,  $(r>r_0, ~0\leq \theta<\pi/2)$, and $(r>r_0, ~\pi/2<\theta\leq \pi)$.
We denote each region by $I$, $N$, and $S$, respectively (Fig. \ref{fig:regions}).}
We look for the set of parameters that satisfy \eref{fromNpole} {in $N$} and \eref{fromSpole} {in $S$.}
Since these are four equations among eight unknown parameters,
the remaining parameters we have to determine are four.

As in the case of the contribution of the angular momentum perturbation, 
\eref{eq:Psiangmom}, we add $\Psi_{\rm H}$ only at $r>r_0$.
Here, we note the symmetry of $\Psi_{\rm P}$. 
From \eref{PsiP}, we find that, 
just like $\psi_0$ and $\psi_4$, the real and imaginary part of $\Psi_{\rm P}$ 
are symmetric and antisymmetric about the equatorial plane respectively. 
In order to kill the jump of  $\Psi_{\rm P}$ at $r=r_0$,  $\Psi_{\rm H}$ at $r>r_0$
must have the same symmetry about the equatorial plane. 
Therefore we get
\begin{equation}
\begin{split}
&
a_{N}(r) = -\overline{a_{S}}(r)~,~~~~~b_{N}(r) = \overline{b_{S}}(r)~,\\
&
c_{N}(r) = -\overline{c_{S}}(r)~,~~~~~d_{N}(r) = \overline{d_{S}}(r)~.
\label{NSrelation}
\end{split}
\end{equation}
Here, $a_{N}(r)$ means $a(r)$ in $N$, and $a_{S}(r)$ means $a(r)$ in $S$, etc. 
It is sufficient if we determine four complex parameters 
only in the region $N$ or $S$.
From \eref{fromNpole}, we adopt $a_1$, $a_2$, $b_1$ and $b_2$ of $\Psi_{\rm H}$ in region $N$ 
as independent parameters. 
When the parameters satisfy \eref{fromNpole}, 
the fields corresponding to $\Psi_{\rm H}$ and $\Psi_{\rm H}$ in $N$ can be written 
as they include only $a_1$, $a_2$, $b_1$ and $b_2$ (equations \eref{psiH}-\eref{HertzH}).

We numerically determine values of these parameters that satisfy the {continuity} conditions
\begin{equation*}
\begin{split}
\left[ F_{\rm P}(r,\theta) \right]_{r_0} + F_{\rm H}(r_0, \theta) = 0~
\end{split}
\end{equation*}
for $F=\psi_1,~\psi_2,~\psi_3,~h_{22},~h_{23},~h_{33},~\Psi$, where
\begin{equation}
\left[ F_{\rm P}(r,\theta)\right]_{r_0} 
\equiv \lim_{r\rightarrow r_0{}^+}F_{\rm P}(r,\theta) - \lim_{r\rightarrow r_0{}^-} F_{\rm P}(r,\theta)~.
\end{equation}
By using the relations between these four parameters, $a_1$, $a_2$, $b_1$ and $b_2$,
with $F_{\rm H}$ above given in \eref{psiH}-\eref{HertzH}, we obtain
\begin{equation*}
\begin{split}
& (a_1)_N = -0.0000025233 - 4.2486\im~,\\
& (a_2)_N = -134.33 - 2123.8\im~,\\
& (b_1)_N = 67.169 + 34.993\im~,\\
& (b_2)_N = -738.86 - 0.079440\im~.\\
\end{split}
\end{equation*}
when $M=1~,m=M/100~,r_0=10M$.
\begin{figure*}[htbp]
\begin{center}
\includegraphics[width=5.6cm]{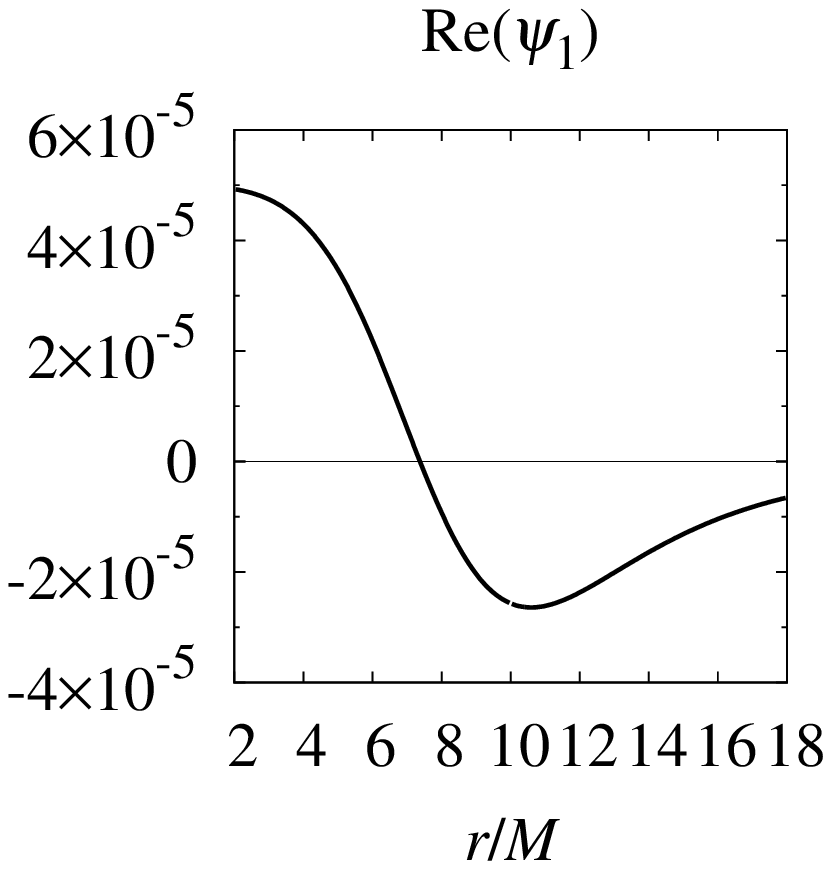}
\includegraphics[width=5.6cm]{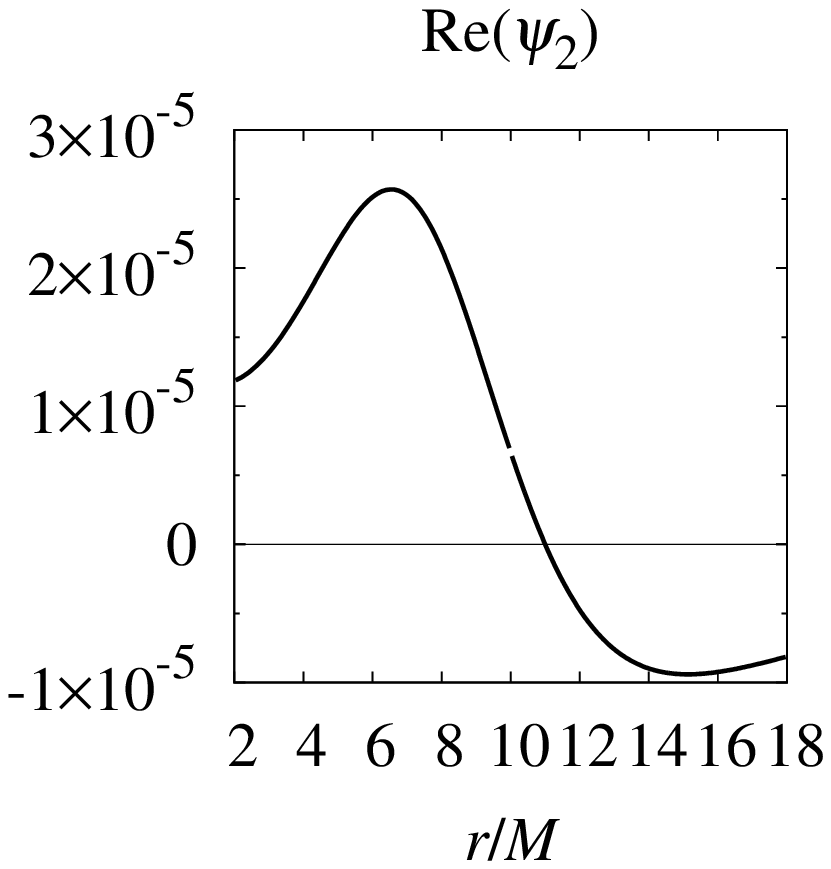}
\includegraphics[width=5.6cm]{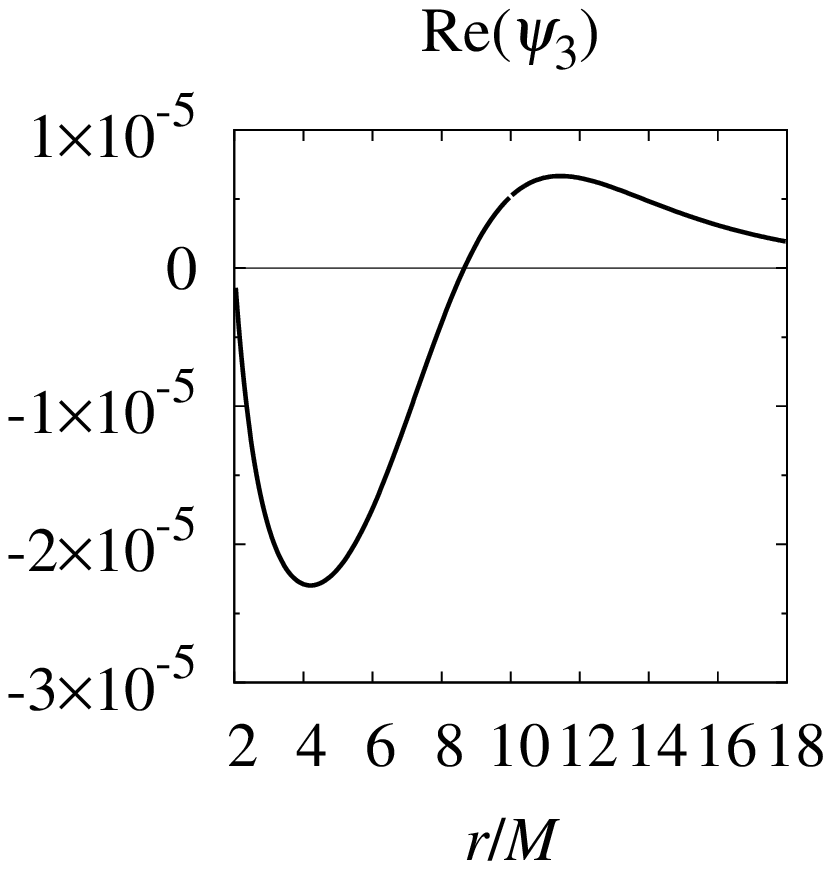}
\end{center}
\caption{Radial dependence of the real part of $\psi_1$ (left), $\psi_2$ (center), and $\psi_3$ (right) 
derived from  $\Psi_{\rm P}+\Psi_{\rm H}$, with $\theta=\pi/4$ fixed. 
The radius of the ring is $r_0=10M$. {It is clear that they are continuous at $r=r_0$.}}
\label{fig:Repsi123}
\end{figure*}
\begin{figure*}[htbp]
\begin{center}
\includegraphics[width=5.6cm]{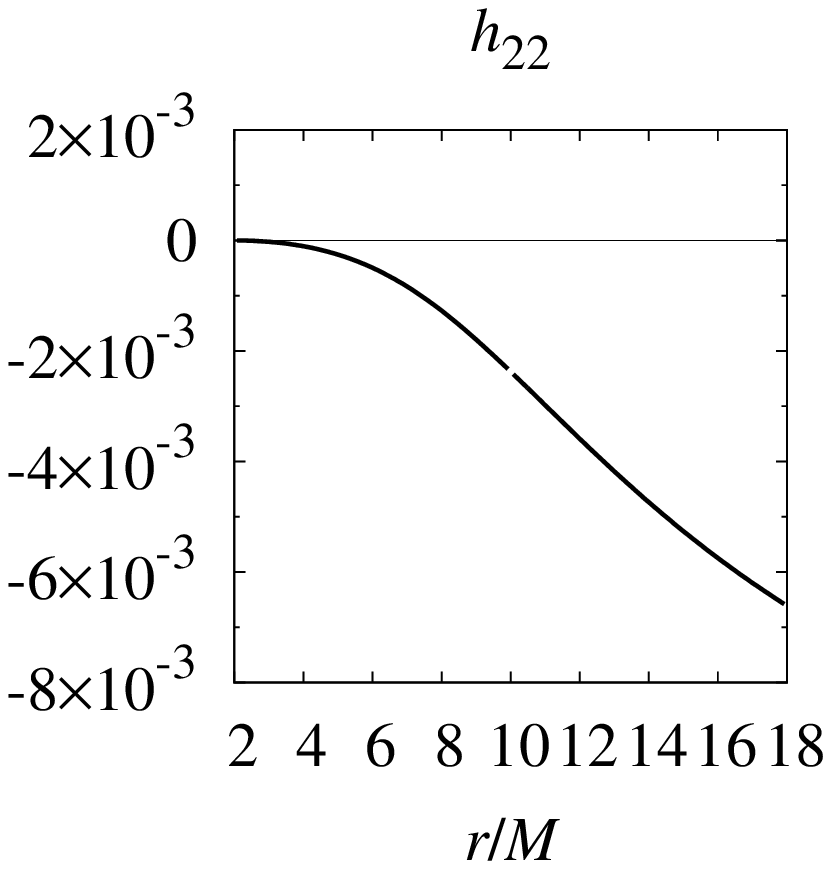}
\includegraphics[width=5.6cm]{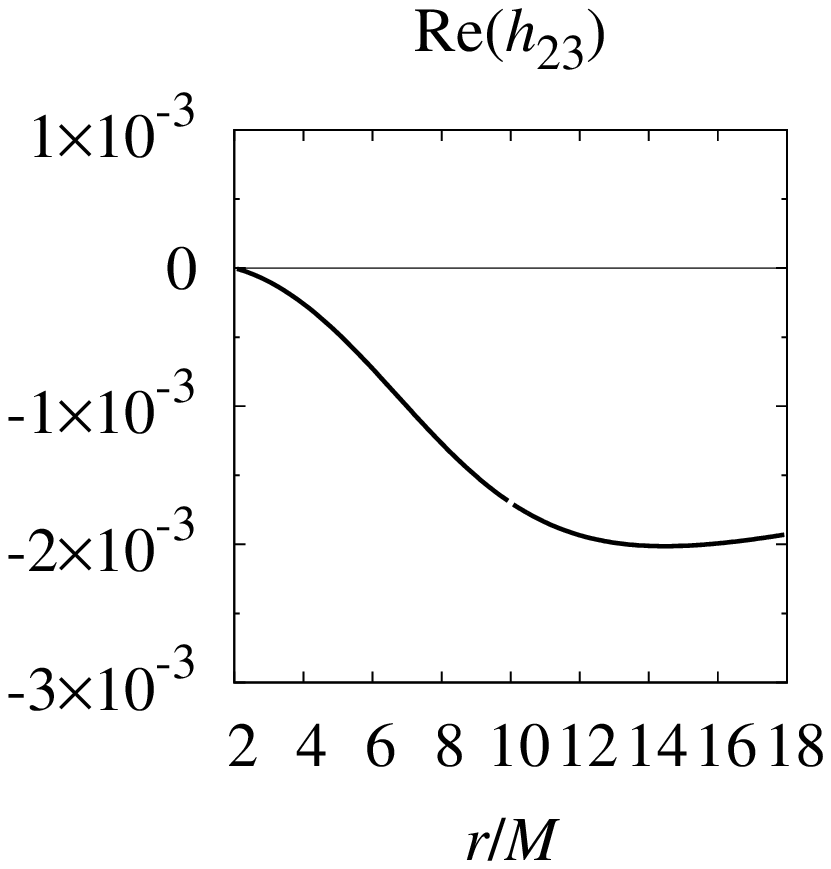}
\includegraphics[width=5.6cm]{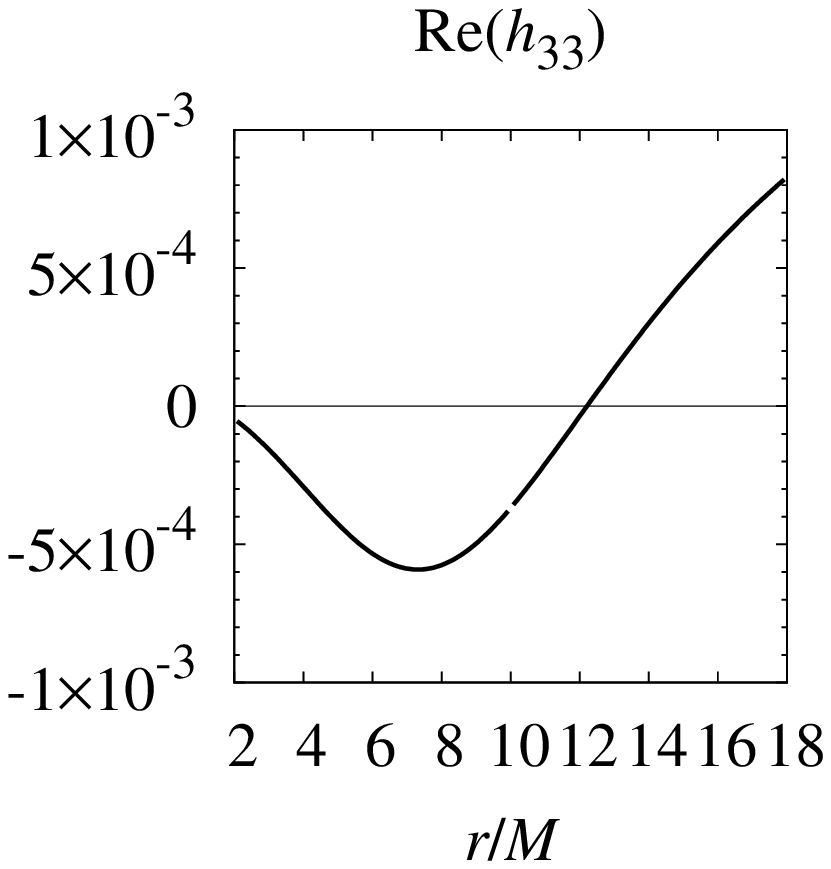}
\\
\includegraphics[width=5.6cm]{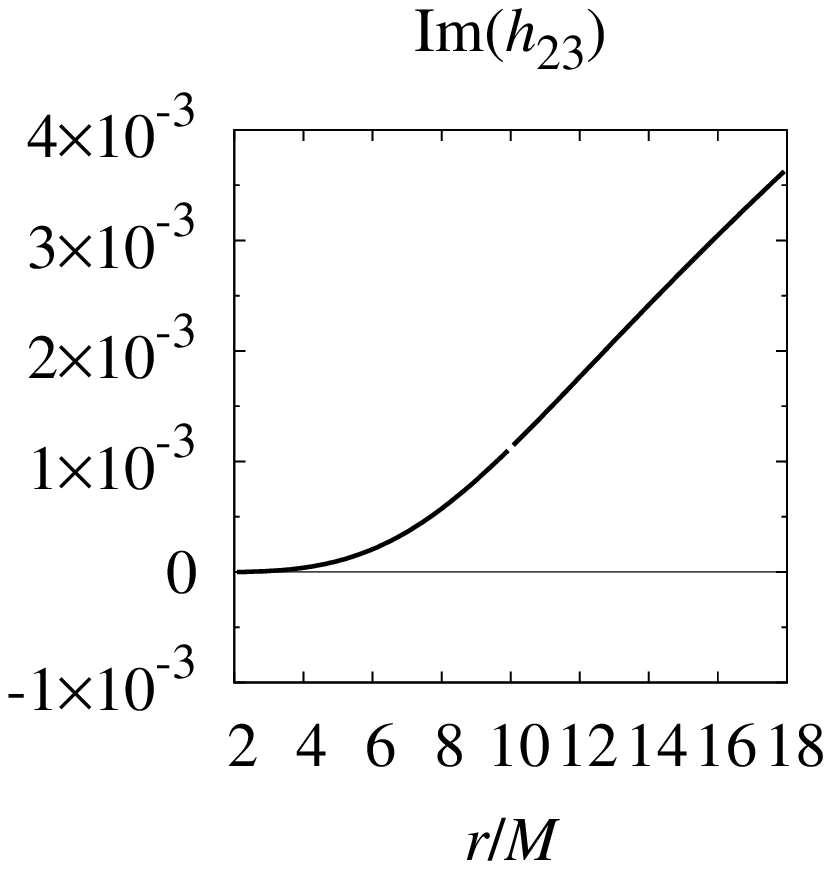}
\includegraphics[width=5.6cm]{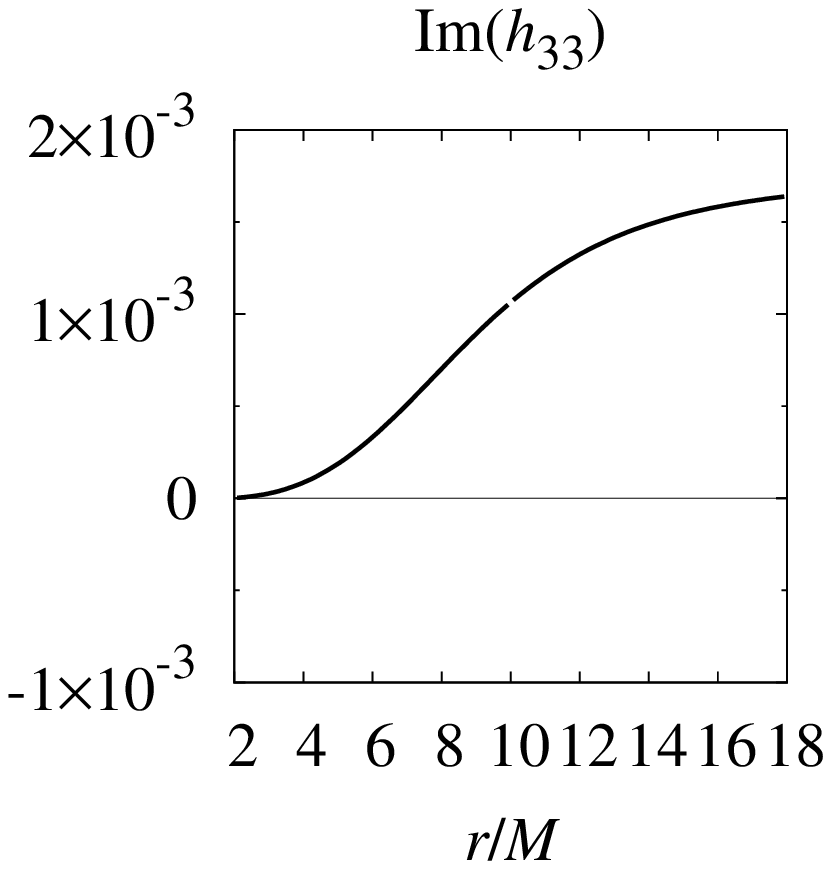}
\end{center}
\caption{
Radial dependence of the each component of $h_{ab}$ derived from $\Psi$ at $\theta=\pi/4$. 
The radius of the ring is $r_0=10M$. They are continuous at $r=r_0$.  }
\label{fig:MP}
\end{figure*}
The plots of $\Re(\psi_1)$, $\Re(\psi_2)$ and $\Re(\psi_3)$ 
derived from $\Psi_{\rm P}+\Psi_{\rm H}$ are shown in Fig. \ref{fig:Repsi123}. 
We find that all of the discontinuity disappeared. 
Note that because of the relations \eref{NSrelation}, 
each of parameters $\Re(b_1)$, $\Re(b_2)$, and $\Im(a_2)$ is the same value
in $N$ and $S$. 
Thus, $\delta M$ and $\delta J$ in \eref{deltaMdeltaJ} is the same in $N$ and $S$. 
Interestingly, we numerically obtain the very good agreement between ($\delta M$, $\delta J$)
and the mass and angular momentum of the ring, \eref{MJring}. 
We obtain from \eref{deltaMdeltaJ},
\begin{equation}
\begin{split}
\delta M &= -A(3M\Re(b_1)+\Re(b_2)) = {0.0600781}
~{,}\\
\delta J &= -A\Im(a_2) = {0.237451}
~.\\
\end{split}
\end{equation}
On the other hand, from \eref{MJring}
\begin{equation}
\begin{split}
M_{\rm ring} &= {0.06007874270}
~,\\
J_{\rm ring} &= {0.2374820823}
~{.}\\
\end{split}
\end{equation}
Although the method to determine the $\Psi_{\rm H}$ here is rather heuristic,
this excellent agreement suggests the validity of the method and the results. 
{Further discussion on the the accuracy of the numerical results is given 
at the end of Appendix \ref{section:DeterminationPsiH}.}

The results in the case of $r_0/M=6, 10, 20, 50$ are shown in Table \ref{r0depM} and \ref{r0depJ}.
\begin{table}[htb]
  \caption{$\delta M$}
  \begin{tabular}{| c | | l | l | l |} \hline
$r_0/M$ & 
$\delta M$ & $M_{\rm ring}$ & $|(M_{\rm ring}-\delta M)/M_{\rm ring}|$  \\ \hline
6  & 0.0592444 & 0.05923843916 & 1.008027909 $\times 10^{-4}$ \\
10 & 0.0600781 & 0.06007874270 & 1.005730101 $\times 10^{-5}$ \\
20 & 0.0613351 & 0.06133564195 & 8.821135362 $\times 10^{-6}$ \\
50 & 0.0622144 & 0.06221386387 & 7.995806223 $\times 10^{-6}$ \\ 
100 & 0.0625205 & 0.06252015946 & 5.948001469 $\times 10^{-6}$ \\
\hline
  \end{tabular}
\label{r0depM}
\end{table}

\begin{table}[htb]
  \caption{$\delta J$}
  \begin{tabular}{| c | | l | l | l |} \hline
$r_0/M$
& $\delta J$ & $J_{\rm ring}$ & $|(J_{\rm ring}-\delta J)/J_{\rm ring}|$ \\ \hline
6  & 0.217649 & 0.2176559237 & 3.301954698 $\times 10^{-5}$ \\
10 & 0.237451 & 0.2374820823 & 1.308149216 $\times 10^{-4}$ \\
20 & 0.304774 & 0.3047792551 & 1.758912364 $\times 10^{-5}$ \\
50 & 0.458263 & 0.4582483860 & 3.190540426 $\times 10^{-5}$ \\ 
100 & 0.637962 & 0.6379608107 & 1.972221458 $\times 10^{-6}$ \\
\hline
  \end{tabular}
\label{r0depJ}
\end{table}
Finally, we show the radial dependence of the metric perturbation,
$h_{22}, \Re(h_{23}), \Re(h_{33}), \Im(h_{23})$, and $\Im(h_{33})$, 
computed from \eref{irg} in Fig. \ref{fig:MP}.
These are the cases for $\theta=\pi/4$. 
We find that they are smooth at $r=r_0$.

\section{Summary and Discussion}
\label{section:summary}
We computed the metric perturbation produced by a rotating circular mass ring 
around a Schwarzschild black hole by using {the} CCK formalism. 
In {the} CCK formalism, {the} Weyl scalars and the metric perturbation are expressed 
by the Hertz potential in a radiation gauge. 
The Hertz potential can be obtained by integrating an equation
which relates the Hertz potential with the Weyl scalars $\psi_0$ or $\psi_4$. 
We used $\psi_4$ to obtain the Hertz potential. 
The Hertz potential contains two parts, $\Psi_{\rm P}$ and $\Psi_{\rm H}$.
$\Psi_{\rm P}$ is derived directly from $\psi_4$ and 
$\Psi_{\rm H}$ is the part which contains the integration constants. 

We first obtained $\Psi_{\rm P}$  
which has discontinuity on the surface of the sphere
at the radius of the ring. $\Psi_{\rm H}$, on the other hand, has 8 complex parameters,
given in \eref{abcd}. Among them, $\Im(a_2)$ is related to the angular momentum 
perturbation and $\Re(b_1)$ and $\Re(b_2)$ are related to the mass perturbation. 
We found that if we determine $\Im(a_2)$ by setting the angular momentum perturbation equal to the 
angular momentum of the ring, the imaginary parts of $\psi_1$, $\psi_2$ and $\psi_3$ 
become continuous at the radius of the ring. 

We determined other parameters by requiring the continuity condition at the radius of the ring. 
We found that if we require the {regularity} condition both at $\theta=0$ and $\theta=\pi$,
we only have a trivial solution and $\Psi_{\rm H}$ becomes zero. 
This fact shows the impossibility to obtain a globally regular solution 
which were discussed previously (\cite{ori03}, \cite{kfw}, \cite{pmb}).
We divided the space time into 3 regions, $N$, $S$ and $I$, as in 
Fig. \ref{fig:regions}, and tried to obtain a solution which is regular in each region and continuous 
on the surface of the sphere at the ring radius. 
We set $\Psi_{\rm H}=0$ in the inner region $I$, and 
determined all unknown parameters of $\Psi_{\rm H}$ 
in the region $N$ and $S$ numerically
by requiring the continuity at the ring radius. 
As a result, the Weyl scalars, $\psi_1$, $\psi_2$ and $\psi_3$, and 
the components of the metric perturbation $h_{\mu\nu}$ become continuous 
at the ring radius. 
We also found that the mass perturbation determined in this method 
agreed with the mass of the ring. 
This fact suggests the validity of the method and the results in this paper. 

The metric perturbation we obtained has a discontinuity on the equatorial plane outside the ring. 
This is similar to the metric perturbation of a Schwarzschild black hole by a particle at rest, 
which was discussed by Keidl et al. \cite{kfw}
Their metric perturbation has radial string singularity inside or outside the particle. 
One of the major difference between Ref. \cite{kfw} and this paper is the presence of the 
angular momentum perturbation in this paper. 
We found that the angular momentum perturbation was important to remove
the discontinuity of $\Im(\psi_1^{\rm P})$, $\Im(\psi_2^{\rm P})$, 
and $\Im(\psi_3^{\rm P})$. 
However, in order to remove the discontinuity 
of the real part of {the} Weyl scalars and that of the metric perturbation,
the mass perturbation $M_{\rm ring}$ and the gauge freedom 
must be added outside the ring. 

A natural extension of this work is to apply to the Kerr black hole case. 
{In the case of Schwarzschild black hole, 
the radial functions $R_l^{(2)}$ and  $R_l^{(-2)}$ were expressed in terms of the associated Legendre functions. 
In the case of Kerr, 
{the radial functions become more complicated.}
Further, the relations between the perturbed Weyl scalars and the Hertz potential 
{become} more complicated. 
{Besides these complication, 
it would be useful to derive the relation between the parameters in $\Psi_{\rm H}$ 
and the mass and angular momentum perturbation in the Kerr case. } 
}

{Will \cite{will74, will75} derived a solution of rotating mass ring around a slowly rotating black hole. 
The method used in those papers are completely different from our method.
Further, the gauge condition used is different from ours. 
We have to treat these issues to compare our results with \cite{will74, will75},
and this is also one of our future works. }

An another interesting and important problem is the case of a particle 
orbiting around a black hole. (e.g., Ref. \cite{pmb})
{In that case, since the problem becomes non-stationary, 
the Teukolsky equation and the spin-weighted spheroidal harmonics
must be solved numerically. Although the problem must be solved fully numerically, 
it would be straightforward to obtain the gravitational field produced by a orbiting particle
by using the method in this paper. }
{Pound et al. \cite{pmb} discussed a method to compute the gravitational self-force 
on a orbiting point mass in {a} radiation gauge by using a local gauge transformation.
Once we obtain the gravitational field in {a} radiation gauge,
it would be possible to compute the self-force with the prescription of \cite{pmb}.}

We will work on these problem in the future.

%


\appendix
\renewcommand\theequation{\Alph{section}.\arabic{equation}}
\section{Newman--Penrose formalism and Teukolsky equation}
\label{section:NP}
{In this appendix, we describe the definition of the Newman--Penrose 
variables, the Teukolsky equation, and the spin weighted spherical harmonics,
which are used in this paper.  
We assume the background Schwarzschild metric is given by (\ref{eq:schwametric}).
}

The null tetrad used in the Newman--Penrose formalism,
\begin{eqnarray}
&&(e_1)^a \equiv l^a = \frac{r^2}{\Delta}(\partial_t)^a+(\partial_r)^a~,\\
&&(e_2)^a \equiv n^a = \frac{1}{2}\left( (\partial_t)^a-\frac{\Delta}{r^2}(\partial_r)^a \right)~,\\
&&(e_3)^a \equiv m^a = \frac{1}{\sqrt{2}r}\left( (\partial_\theta)^a+\im\csc\theta(\partial_\phi)^a \right)~,\\
&&(e_4)^a \equiv \overline{m}^a = \frac{1}{\sqrt{2}r}\left( (\partial_\theta)^a -\im\csc\theta(\partial_\phi)^a \right)
\end{eqnarray}
satisfies normalization and orthogonality conditions.
\begin{equation}
\begin{split}
& l_a l^a = n_a n^a = m_a m^a = \overline{m}_a \overline{m}^a =0~,\\
& l_a m^a = l_a\overline{m}^a = n_a m^a = n_a\overline{m}^a =0~,\\
& -l_a n^a = m_a \overline{m}^a =1~.
\end{split}
\end{equation}
The coordinate basis is denoted by $(\partial_\mu)^a$. We define directional derivatives,
\begin{equation}
\begin{split}
& {\pmb D} = l^a\partial_a = {}_{,1}~,
~~~~~
{\pmb \Delta} = n^a\partial_a = {}_{,2}~,\\
& {\pmb \delta} = m^a\partial_a = {}_{,3}~,
~~~~~
\overline{\pmb \delta} = \overline{m}^a\partial_a = {}_{,4}~,
\end{split}
\end{equation}
where $\partial_a$ is ordinary derivative associated with the coordinate basis. We also use auxiliary symbols $\tilde{\pmb{D}}$ and $\tilde{\pmb\Delta}$.
\begin{equation}
\tilde{\pmb{D}}\equiv \left(  -\frac{2r^2}{\Delta} \right) \pmb{\Delta}~,
~~~~~
\tilde{\pmb\Delta} \equiv \left( -\frac{\Delta}{2r^2} \right) \pmb{D}~.
\end{equation}

{The} Ricci rotation coefficients $\gamma_{\mu\nu\rho}$ are defined as
\begin{equation}
\gamma_{\mu\nu\rho} \equiv (e_\mu)_{a;b} (e_\nu)^a (e_\rho)^b
~,
\label{ricci}
\end{equation}
where ``$_;$'' 
represents covariant derivative. 
Nonzero components of $\gamma_{\mu\nu\rho}$ becomes 
\begin{equation}
\begin{split}
&
\gamma_{122} = -\gamma_{212} = -\frac{M}{r^2} = -2\gamma~,\\
&
\gamma_{134} = -\gamma_{314} = \gamma_{143} = -\gamma_{413} = \frac{1}{r} = -\rho~,\\
&
\gamma_{234} = -\gamma_{324} = \gamma_{243} = -\gamma_{423} = -\frac{\Delta}{2r^3} = \mu
~,\\
&
\gamma_{343} = -\gamma_{433} = \gamma_{434} = -\gamma_{344}  = \frac{\cot\theta}{\sqrt{2}r} = 2\beta~.
\end{split}
\end{equation}
%

The master perturbation equation is written as
\begin{equation}
L_{(s)} \psi_{(s)} = 4\pi {T_{(s)}}~,
\end{equation}
where
\begin{equation}
\begin{split}
L_{(s)} &\equiv 
\frac{r^2}{\Delta}\partial_t{}^2
-2s\left(\frac{M}{\Delta}-\frac{1}{r}\right)\partial_t  
-\frac{\Delta^{-s}}{r^2}\partial_r \left(\Delta^{s+1}\partial_r \right)
\\&~~~~~
-\frac{1}{r^2} \Big[ 
\csc\theta\partial_\theta(\sin\theta\partial_\theta) - s^2\cot^2\theta + s
\\&~~~~~~~~~~
+ 2s\im \csc^2\theta\cos\theta\partial_\phi + \csc^2\theta\partial_\phi{}^2
\Big]
~.
\end{split}
\end{equation}
Putting $s=2$ or $s=-2$, the equation becomes an equation 
for $\psi_0$ and $\psi_4$, respectively. 
\begin{eqnarray}
\psi_{(s=2)} &= \psi_0~,\quad\quad 
\psi_{(s=-2)} &= \rho^{-4}\psi_4
~{.}
\end{eqnarray}
The source term becomes for $s=2, -2$, 
\begin{equation}
\begin{split}
& {T_{(s=2)}} = 
-2({\pmb \delta}-2\beta){\pmb \delta}T_{11}
\\&~~~~~~~~~~~~~~~
+4({\pmb D}-4\rho)({\pmb \delta}-2\beta)T_{13}
\\&~~~~~~~~~~~~~~~~~~~~
-2({\pmb D}-5\rho)({\pmb D}-\rho)T_{33}
~{,}\\
\end{split}
\end{equation}
\begin{equation}
\begin{split}
\rho^{4} &{T_{(s=-2)}} = 
-2(\overline{\pmb\delta}-2\beta)\overline{\pmb\delta} T_{22}
\\&~~~~~~~~~~~~~~~
+4(\pmb\Delta+4\mu+2\gamma)(\overline{\pmb\delta}-2\beta)T_{24}
\\&~~~~~~~~~~~~~~~~~~~~
-2(\pmb\Delta+5\mu+2\gamma)(\pmb\Delta+\mu)T_{44}~,
\end{split}
\end{equation}
where $T_{\mu\nu} = T_{ab}(e_{\mu})^a (e_{\nu})^b$. The source term 
{$T_{(s=-2)}$} 
can also be expressed as
\begin{equation}
\begin{split}
\frac{4}{\Delta^2} {T_{(s=-2)}} &= 
-2 (\overline{\pmb\delta}-2\beta)\overline{\pmb\delta} \frac{4r^4}{\Delta^2} T_{22}
\\&~~~~~
-4(\tilde{\pmb D}-4\rho)(\overline{\pmb\delta}-2\beta) \frac{2r^2}{\Delta} T_{24}
\\&~~~~~~~~~~
-2(\tilde{\pmb D}-5\rho)(\tilde{\pmb D}-\rho)T_{44} ~.
\end{split}
\end{equation}
In this expression we see the symmetry between $T_{(s=2)}$ and $T_{(s=-2)}$.

The equation can be separated as 
\begin{equation}
\psi_{(s)} 
= \sum_{l,m}^{\infty} \sekibun{-\infty}{\infty}{\omega} 
R^{(s)}_{lm\omega}(r)~_sY_{lm}(\theta, \phi)\ex{-\im \omega t}~,
\end{equation}
where $_sY_{lm}(\theta, \phi)$ is spin-weighted spherical harmonics. 
Equations for radial and angular part are
\begin{equation}
\begin{split}
&
\Delta^{-s} \bibun{}{r}{}\left( \Delta^{s+1} \bibun{}{r}{} \right) R^{(s)}_{lm\omega}
\\&~~~~~
+\left[ \frac{r^4\omega^2 - 2\im s(r-M)r^2\omega}{\Delta} + 4\im s\omega r \right] R^{(s)}_{lm\omega}
\\&~~~~~~~~~~
-(l-s)(l+s+1)R^{(s)}_{lm\omega} = {-} 4\pi r^2 T^{(s)}_{lm\omega}~, 
\label{bunriR}
\end{split}
\end{equation}
\begin{equation}
\begin{split}
& \left[ \csc\theta\partial_\theta(\sin\theta\partial_\theta) - s^2\cot^2\theta + s \right]{}_sY_{lm}
\\&~~~~~
+ \left( 2s\im \csc^2\theta\cos\theta\partial_\phi + \csc^2\theta\partial_\phi{}^2  \right){}_sY_{lm}
\\&~~~~~~~~~~
+(l-s)(l+s+1)~{}_sY_{lm}=0~.
\label{bunriY}
\end{split}
\end{equation}
This separated equation \eref{bunriR} is called the Teukolsky equation. 
The source term $T^{(s)}_{lm\omega}$ is defined as 
\begin{equation}
T^{(s)}_{lm\omega}
= \sekibun{-\infty}{\infty}{t}
\sekibun{0}{\pi}{\theta}
\sekibun{0}{2\pi}{\phi}
\sin\theta
~_s\overline{Y}_{lm}(\theta, \phi)\ex{\im\omega t} {T_{(s)}}~.
\end{equation}

The angular part \eref{bunriY} is the eigen value equation for $_sY_{lm}(\theta, \phi)$. 
The spin-weighted spherical harmonics is defined as
\[
_sY_{lm} = \left\{ \begin{array}{ll}
    \sqrt{\frac{(l-s)!}{(l+s)!}}~\eth^s{}Y_{lm}(\theta,\phi) &(0\leq s\leq l)~, \\
    (-1)^s\sqrt{\frac{(l+s)!}{(l-s)!}}~\overline{\eth}^{-s}{}Y_{lm}(\theta,\phi) &(-l\leq s\leq 0)~,
  \end{array} \right.
\]
where $Y_{lm}~(={}_0Y_{lm})$ is ordinal spherical harmonics, 
and $\eth$ {and $\overline{\eth}$ are} partial derivative operators defined as
\begin{eqnarray}
&&\eth~{}_sY_{lm} = -\left( \partial_\theta+\im\csc\theta\partial_\phi-s\cot\theta \right) {}_sY_{lm}~,\\
&&\overline{\eth}~{}_sY_{lm} = -\left( \partial_\theta-\im\csc\theta\partial_\phi+s\cot\theta \right) {}_sY_{lm}~.
\end{eqnarray}
For a fixed value of $s$ of the spin weight, the set of the spin-weighted spherical harmonics is complete and orthonormal.
\begin{equation}
\begin{split}
\sum_{l=|s|}^{\infty}\sum_{m=-l}^{l} & {}_s\overline{Y}_{lm}(\theta',\phi')~_sY_{lm}(\theta,\phi)
\\&~~~~~
=\frac{1}{\sin\theta}\delta(\theta-\theta')\delta(\phi-\phi')
~,
\end{split}
\end{equation}
\begin{equation}
\begin{split}
\sekibun{0}{\pi}{\theta}\sekibun{0}{2\pi}{\phi} & ~\sin\theta~{}_s\overline{Y}_{lm}(\theta,\phi)~{}_sY_{l'm'}(\theta,\phi)
\\&~~~~~
=\delta_{ll'}\delta_{mm'}
~.
\end{split}
\end{equation}
For a fixed value of $s$, any function of $(\theta, \phi)$ with spin weight $s$ can be 
expanded by $_sY_{lm}(\theta,\phi)$ \cite{np66, castillo}.

By definition, the differential operator $\eth~ (\overline{\eth})$ raises (lowers) the spin weight $s$ of the spin weighted spherical harmonics.
\begin{eqnarray}
\eth~{}_sY_{lm} = +\sqrt{(l-s)(l+s+1)}~{}_{s+1}Y_{lm}~,
\label{spinup}\\
\overline{\eth}~{}_sY_{lm} = -\sqrt{(l+s)(l-s+1)}~{}_{s-1}Y_{lm}~.
\label{spindown}
\\
\eth\overline{\eth}~{}_sY_{lm} =- (l+s)(l-s+1)~{}_sY_{lm}~,
\\
\overline{\eth}\eth~{}_sY_{lm} = -(l-s)(l+s+1)~{}_sY_{lm}~.
\label{ethbareth}
\end{eqnarray}
The angular part of the perturbation equation \eref{bunriY} is identical to the equation \eref{ethbareth}. 
The four equations \eref{spinup} to \eref{ethbareth} can be rewritten using notation from {the} Newman--Penrose formalism.
\begin{equation}
\begin{split}
&
(\pmb\delta -2s\beta){}_sY_{lm} 
= +\frac{\rho}{\sqrt{2}}\sqrt{(l-s)(l+s+1)}~{}_{s+1}Y_{lm} ~,\\
&
(\overline{\pmb\delta} +2s\beta){}_sY_{lm} 
= -\frac{\rho}{\sqrt{2}}\sqrt{(l+s)(l-s+1)}~{}_{s-1}Y_{lm} ~.\\
&
(\overline{\pmb\delta} +2(s+1)\beta)(\pmb\delta -2s\beta){}_sY_{lm} \\
&~~~~~~~~~~~~~~~~~~~~ = -\frac{\rho^2}{2}(l-s)(l+s+1){}_sY_{lm}~,\\
&
(\pmb\delta -2(s-1)\beta)(\overline{\pmb\delta} +2s\beta){}_sY_{lm} \\
&~~~~~~~~~~~~~~~~~~~~ = -\frac{\rho^2}{2}(l+s)(l-s+1){}_sY_{lm}~.
\end{split}
\end{equation}

Following relation also hold{s}.
\begin{equation}
_s\overline{Y}_{lm}(\theta, \phi) = (-1)^{m+s}{}_{-s}Y_{lm}(\theta, \phi)~.
\end{equation}
%

\section{Solutions of the Teukolsky equation}
\label{section:solTeukolsky}
In this appendix, we explain how to derive solutions of the Teukolsky equation,
\eref{eq:TeuR2} and \eref{eq:TeuR2m}.
Each of \eref{dokei0} and \eref{dokei4} is solved by using the Green's function. 
For $\psi_0$, we look for a Green's function $G_l^{(2)}(r,r')$ that satisfies
\begin{equation}
\begin{split}
\left[ \bibun{}{r}{}\left(\Delta^3\bibun{}{r}{} \right) - \Delta^2 (l-2)(l+3) \right] G_l^{(2)} 
= -\delta(r-r')
\label{G2}
\end{split}
\end{equation}
and obtain $R_l^{(2)}(r)$ by
\begin{equation}
R_l^{(2)}(r) = \sekibun{}{}{r'} 
\left[ G_l^{(2)}(r,r') \left( 4\pi T_l^{(2)}(r') r'^2 \Delta'^2 \right) \right] ~,
\end{equation}
where $\Delta'=r'^2-2Mr'$.

For $\psi_4$, we look for a Green's function  $G_l^{(-2)}(r,r')$ that satisfies
\begin{equation}
\begin{split}
\left[ \bibun{}{r}{}\left(\frac{1}{\Delta}\bibun{}{r}{} \right) - \frac{(l+2)(l-1)}{\Delta^2} \right] G_l^{(-2)}
= -\delta(r-r')
\label{G-2}
\end{split}
\end{equation}
and obtain $R_l^{(-2)}(r)$ by
\begin{equation}
R_l^{(-2)}(r) = \sekibun{}{}{r'} 
\left[ G_l^{(-2)}(r,r') \left( 4\pi T_l^{(-2)}(r') \frac{r'^2}{\Delta'^2} \right) \right]~.
\end{equation}

The ``peeling off theorem'' \cite{np62} states that 
the the asymptotic behaviors of the Weyl scalars at $r \rightarrow \infty$ are
\begin{equation}
\psi_0 = \mathcal{O}(r^{-5})~,
~~~~~
\psi_4 = \mathcal{O}(r^{-1})
\end{equation}
without ingoing waves, and 
\begin{equation}
\psi_0 = \mathcal{O}(r^{-1})~,
~~~~~
\psi_4 = \mathcal{O}(r^{-5})
\end{equation}
without outgoing waves. In the case of our problem, 
since there is no radiation, the asymptotic behaviors are
\begin{equation}
\psi_0 = \mathcal{O}(r^{-5})~,
~~~~~
\psi_4 = \mathcal{O}(r^{-5})~.
\end{equation}
Therefore, the asymptotic behaviors of the Green's functions 
and the radial functions are
\begin{eqnarray}
&&
\psi_0 \sim R_l^{(2)} \sim G_l^{(2)} = \mathcal{O}(r^{-5})~,
\\
&&
r^4 \psi_4 \sim R_l^{-(2)} \sim G_l^{(-2)} = \mathcal{O}(r^{-1})~.
\end{eqnarray}

The Green's function is found in a form of
\begin{equation}
\begin{split}
G_l^{(s)}(r,r')
&
= \frac{h_1^{(s)}(r) h_2^{(s)}(r')}{W^{(s)}} \Theta(r'-r)
\\ &~~~~~
+ \frac{h_1^{(s)}(r') h_2^{(s)}(r)}{W^{(s)}} \Theta(r-r')~,
\end{split}
\end{equation}
where ${h_1^{(s)}}$ and ${h_2^{(s)}}$ are independent homogenous solutions of equation \eref{G2} (\eref{G-2}), 
and $W^{(s)}$ is defined as
\begin{equation}
\begin{split}
W^{(2)} &= -\Delta^3 \left[ h_1^{(2)}\bibun{h_2^{(2)}}{r}{} - h_2^{(2)}\bibun{h_1^{(2)}}{r}{} \right] ~,
\\
W^{(-2)} &= -\frac{1}{\Delta} \left[ h_1^{(-2)}\bibun{h_2^{(-2)}}{r}{} - h_2^{(-2)}\bibun{h_1^{(-2)}}{r}{} \right] ~.
\end{split}
\end{equation}

For $\psi_0$,  
\begin{equation}
h_1^{(2)}(r) = \frac{P_l^2(x)}{\Delta}~, 
~~~
h_2^{(2)}(r) = \frac{Q_l^2(x)}{\Delta}~,
\end{equation}
where $P_l^2$ and $Q_l^2$ are associated Legendre functions, and $x\equiv (r-M)/M$, $\Delta=r^2-2Mr=M^2(x^2-1)$. 
For $\psi_4$,  
\begin{equation}
h_1^{(-2)}(r) = \Delta P_l^2(x)~, 
~~~
h_2^{(-2)}(r) = \Delta Q_l^2(x)~.
\end{equation}
Then $W^{(s)}$ becomes
\begin{equation}
W^{(2)} = W^{(-2)} = M\frac{(l+2)!}{(l-2)!} = M(l+2)(l+1)l(l-1)
~.
\end{equation}
Since $h_1|_{r=2M}$ is regular and $h_2|_{r\rightarrow \infty}=0$, each Green's function is regular at the event horizon $r=2M$ and vanishes at infinity and is continuous at $r=r'$.

We write the Green's functions as
\begin{equation}
\begin{split}
&G_l^{(2)}(r,r') = \frac{P_l^2(x'_<) Q_l^2(x'_>)}{M \Delta \Delta' (l+2)(l+1)l(l-1)}~,\\
&G_l^{(-2)}(r,r') = \frac{\Delta \Delta' P_l^2(x'_<) Q_l^2(x'_>)}{M (l+2)(l+1)l(l-1)}~,
\end{split}
\end{equation}
where we define
\begin{equation}
x'_{<}\equiv \frac{{\rm min}(r,r')-M}{M}~~~{\rm and}~~~x'_{>}\equiv \frac{{\rm max}(r,r')-M}{M}~.
\end{equation}

A simple relation $\Delta^2\Delta'^2 G_l^{(2)}(r,r') = G_l^{(-2)}(r,r')$ holds because of symmetries.

\section{{Derivation of Weyl scalar $\psi_3$}}
\label{section:psi3}
In this section, we show a derivation of \eref{eq:psi3Hertz}.
Note that we assume the Schwarzschild metric as a background space-time. 
Some useful identities in the Newman--Penrose formalism used in this section can be found in Ref. \cite{chandra}.

We start from the definition of Weyl scalars \eref{weylscalar-def}.
Since the Weyl tensor is equal to the Riemann curvature tensor at a vacuum point,
the first order perturbation the Weyl tensor, $C^{(1)}_{abcd}$,  can be written as
\begin{equation}
\begin{split}
-2C^{(1)}_{abcd}
& = h_{ac;bd}+h_{bd;ac}-h_{bc;ad}-h_{ad;bc}\\
&~~~~~ +C^{(0)}_{aecd}h^e{}_b-C^{(0)}_{becd}h^e{}_a~,
\label{Ctoh}
\end{split}
\end{equation}
where $C^{(0)}_{abcd}$ is the unperturbed Weyl tensor. 
The nonzero tetrad components of $C^{(0)}_{abcd}$ are 
$C^{(0)}_{1342}=\Psi_2$ and $C^{(0)}_{1212}=C^{(0)}_{3434} 
= -2{\rm Re}(\Psi_2) = -2\Psi_2$.  
The tetrad components of covariant derivative $h_{ab;ef}$ can be written as
\begin{equation}
\begin{split}
h_{\mu\nu;\rho\sigma} & \equiv
h_{ab;ef} (e_\mu)^a (e_\nu)^b (e_\rho)^e (e_\sigma)^f \\
& = [h_{\mu\nu,\rho} + 2h_{\kappa(\mu} \gamma^\kappa{}_{\nu)\rho}]_{,\sigma} \\
&~~~~~ 
+ [h_{\lambda\mu,\rho} 
+2 h_{\kappa(\lambda} \gamma^\kappa{}_{\mu)\rho}] \gamma^\lambda{}_{\nu\sigma}\\
&~~~~~
+ [h_{\lambda\nu,\rho}
+2 h_{\kappa(\lambda} \gamma^\kappa{}_{\nu)\rho}] \gamma^\lambda{}_{\mu\sigma}\\
&~~~~~
+ [h_{\mu\nu,\lambda}
+2 h_{\kappa(\mu} \gamma^\kappa{}_{\nu)\lambda}] \gamma^\lambda{}_{\rho\sigma} ~,
\label{hdiffidff}
\end{split}
\end{equation}
where $\gamma^\lambda{}_{\rho\sigma}$ is the Ricci rotation coefficients \eref{ricci}.

By using \eref{Ctoh} and \eref{hdiffidff}, 
we can obtain an expression for $\psi_3$ in terms of $h_{\mu\nu}$.
\begin{equation}
\begin{split}
-2\psi_3 &= h_{14;22}+h_{22;14}-h_{24;12}-h_{12;24}
+C^{(0)}_{1342}h^3{}_2
\\
&= [\pmb{D}\overline{\pmb \delta} h_{22} -2\mu(\pmb{D}+\rho) h_{24}] 
\\
&~~~~~ - \pmb{\Delta}\pmb{D}h_{24} - (\pmb{\Delta}+2\gamma)\rho h_{24} 
+ 2\gamma\rho h_{24}
\\
&= \pmb{D}\overline{\pmb \delta} h_{22} -(\pmb{\Delta}+2\mu)(\pmb{D}+\rho)h_{24}
~.
\label{psi3eq1}
\end{split}
\end{equation}

By substituting the relation \eref{irg} between $h_{ab}$ and the Hertz potential $\Psi$ into \eref{psi3eq1},
we obtain
\begin{equation}
\begin{split}
-2\psi_3 &= 
-\pmb{D}\overline{\pmb \delta}(\overline{\pmb\delta} +2\beta)(\overline{\pmb\delta} +4\beta)\overline{\Psi} 
-\pmb{D}\overline{\pmb \delta}({\pmb\delta} +2\beta)({\pmb\delta} +4\beta){\Psi}
\\&~~~~~
+(\pmb{\Delta} +2\mu)(\pmb{D} +\rho)(\pmb{D} +\rho)({\pmb\delta} +4\beta){\Psi}~.
\label{psi3eq2}
\end{split}
\end{equation}
The second term of the right hand side of \eref{psi3eq2} becomes 
\begin{equation*}
\begin{split}
- & \pmb{D} \overline{\pmb \delta} ({\pmb\delta} +2\beta)({\pmb\delta} +4\beta){\Psi}
\\
&= 
-[\pmb{\Delta}\pmb{D}\pmb{D}  + 2\pmb{D}\rho\partial_t +6\gamma\pmb{D}\rho]({\pmb \delta} +4\beta){\Psi}
~,
\end{split}
\end{equation*}
where we used the fact the Hertz potential satisfies the source-free Teukolsky equation \eref{IRGhomo1}.
On the other hand, the third term of the right hand side of \eref{psi3eq2} becomes
\begin{equation*}
\begin{split}
(\pmb{\Delta} +2\mu) & (\pmb{D} +\rho)(\pmb{D} +\rho)({\pmb\delta} +4\beta){\Psi}
\\
&=
[\pmb{\Delta}\pmb{D}\pmb{D} +2\pmb{D}\rho\partial_t ]({\pmb\delta} +4\beta){\Psi}~.
\end{split}
\end{equation*}
As a result, the expression for $\psi_3$ in terms of the Hertz potential, Eq. \eref{eq:psi3Hertz} is obtained.
\begin{equation*}
\begin{split}
-2\psi_3
&= -\pmb{D}\overline{\pmb \delta}(\overline{\pmb\delta} +2\beta)(\overline{\pmb\delta} +4\beta)\overline{\Psi} 
-6\gamma\pmb{D}\rho({\pmb\delta} +4\beta){\Psi}~.
\end{split}
\end{equation*}

\section{Determination of all the parameters in $\Psi_{\rm H}$}
\label{section:DeterminationPsiH}
The ``homogeneous solution'' part $\Psi_{\rm H}$ of the Hertz potential has 8 complex parameters. By analyzing its physical contribution to the space-time, $\Im(a_2)$ can be determined analytically.
\begin{equation}
J_{\rm ring} = -A\Im(a_2)~.
\end{equation}
The imaginary parts of all the Weyl scalars are smooth with this value of $\Im(a_2)$. 
However, we do not have analytic formula for other parameters as far as we know.

Thus we determine all the parameters by using the continuity condition on {the} Weyl scalars, 
metric perturbation, and the Hertz potential. Before imposing the continuity condition, 
we reduce the number of parameters as follows. 
Near the poles ($\theta=0,~\pi$), $\Psi_{\rm H}$ is
\begin{equation*}
\begin{split}
\frac{\sin^2\theta}{2A}\overline{\Psi_{\rm H}}
& = -\frac{1}{3}(a_1-3d_1)r^3 + (b_1+c_1-3Md_1)r^2 \\
&~~~~~ -(2M^2a_1-c_2-b_2)(r-M) \\
&~~~~~ {-}\frac{1}{3}(a_2-3d_2)+\frac{M}{2}b_2\\
&~~~~~~~~~~ +\mathcal{O}(\theta^2)~~~{\rm as}~\theta\rightarrow 0~,
\end{split}
\end{equation*}
and
\begin{equation*}
\begin{split}
\frac{\sin^2\theta}{2A}\overline{\Psi_{\rm H}}
& = \frac{1}{3}(a_1+3d_1)r^3 + (b_1-c_1-3Md_1)r^2 \\
&~~~~~ +(2M^2a_1-c_2+b_2)(r-M) \\
&~~~~~ -\frac{1}{3}(a_2+3d_2)+\frac{M}{2}b_2\\
&~~~~~~~~~~ +\mathcal{O}((\pi-\theta)^2)~~~{\rm as}~\pi-\theta \rightarrow 0~.
\end{split}
\end{equation*}
%
On the other hand, 
we see that 
{the} Weyl scalars and metric perturbation corresponding to $\Psi_{\rm P}$ as well as $\Psi_{\rm P}$ 
do not have $\mathcal{O}(\theta^{-1})$ or $\mathcal{O}(\theta^{-2})$ 
behaviors 
as $\theta\rightarrow 0$ and $\pi-\theta \rightarrow $0. 
Therefore, the conditions \eref{fromNpole} and \eref{fromSpole} follow. 

When the parameters satisfy \eref{fromNpole}, the fields 
corresponding to $\Psi_{\rm H}$ and $\Psi_{\rm H}$ in region $N$ ($r>r_0,~0 \leq \theta \leq \pi/2$) can be written as
\begin{equation}
\begin{split}
&
\psi_1^{\rm H} = \frac{ 3 A}{\sqrt{2} r^4} \left[ -a_2\sin\theta + 2M \frac{ b_2(1-\cos\theta) }{\sin\theta} \right]~,\\
&
\psi_2^{\rm H} = \frac{A}{r^4} \left[ (r-3M)b_2 + 3a_2\cos\theta \right]~,\\
&
\psi_3^{\rm H} = \frac{3AM}{\sqrt{2} r^4} \bigg[ \frac{1}{2M} \left( a_2-{2Mr^2}\Re(a_1) \right) \sin\theta
\\
&~~~~~~~~~~~~~~~~~~~~ 
+  \frac{r-2M}{r} \frac{ \overline{b_2}(1-\cos\theta) }{\sin\theta} \bigg]~,
\label{psiH}
\end{split}
\end{equation}
\begin{equation}
\begin{split}
&
h_{22}^{\rm H} = \frac{2A}{r^2} \Big\{ -\left[ r^2\Re(b_1)+(r-M)\Re(b_2) \right]
\\
&~~~~~~~~~~
-\left[ r^2(r-3M)\Re(a_1) + \Re(a_2) \right] \cos\theta \Big\}~,\\
&
h_{23}^{\rm H} = \frac{\sqrt{2}A}{2r^2} \bigg[ -(r^3a_1 - 2a_2)\sin\theta 
\\
&~~~~~~~~~~
+2(r-2M) \frac{ b_2(1-\cos\theta) }{\sin\theta} \bigg]~,
\\
&
h_{33}^{\rm H} = 2A \bigg[   - Ma_1(1-\cos\theta)
\\
&~~~~~~~~~~  + \left( b_1+\frac{b_2}{r}  -Ma_1 \right)\left( \frac{1-\cos\theta}{\sin\theta} \right)^2  \bigg]~,
\label{hH}
\end{split}
\end{equation}
\begin{equation}
\begin{split}
\overline{\Psi_{\rm H}} 
&= \frac{2A}{\sin^2\theta} \bigg[ -\frac{a(r)}{6}(1-\cos^3\theta) -\frac{b(r)}{2}(1-\cos^2\theta) 
\\
&~~~~~~~~~~~~~~~ +\left(\frac{a(r)}{2}+b(r)\right)(1-\cos\theta) \bigg]~.
\label{HertzH}
\end{split}
\end{equation}
%

%
\begin{figure}[ht]
\begin{center}
\includegraphics[width=4cm]{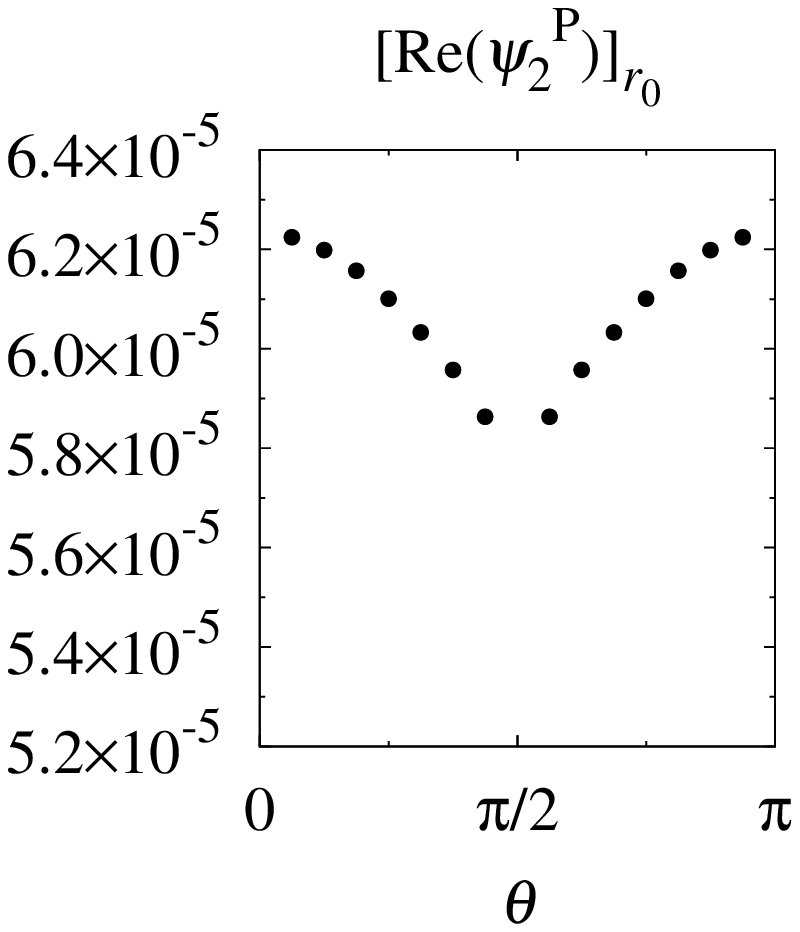}
\includegraphics[width=4cm]{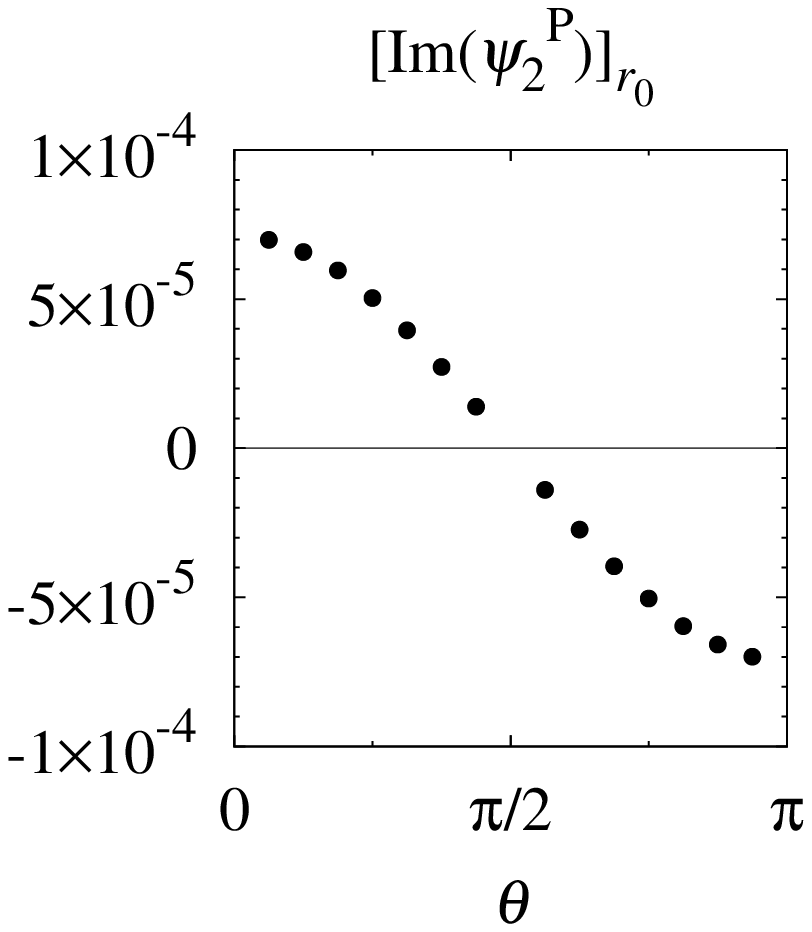}
\end{center}
\caption{Angular dependenc{e} of the jump of $\psi_2^{\rm P}$ at $r=r_0$. The left panel is the real part and the right panel is the imaginary part of $[\psi_2^{\rm P}(r,\theta)]_{r_0}$.}
\label{fig:jump}
\end{figure}
The jumps of fields corresponding to $\Psi_{\rm P}$ depends on $\theta$. 
The plots of the jump of $\psi_2^{\rm P}$ at $r=r_0$, 
$\left[ \psi_2^{\rm P}(r,\theta)\right]_{{r_0}}$
are shown in Fig. \ref{fig:jump} for examples. An extrapolation with a forth order polynomial 
is used to evaluate $\left[ \psi_2^{\rm P}(r,\theta)\right]_{r_0}$.

We can solve
\begin{equation*}
\begin{split}
& \left[ \psi_1^{\rm P}(r,\theta)\right]_{r_0} + \psi_1^{\rm H}(r_0, \theta) = 0~,
\\
& \left[ \psi_2^{\rm P}(r,\theta)\right]_{r_0} + \psi_2^{\rm H}(r_0, \theta) = 0
\end{split}
\end{equation*}
for an arbitrary fixed $\theta$ to obtain $a_2$ and $b_2$. Then we can solve
\begin{equation*}
\begin{split}
& \left[ h_{33}^{\rm P}(r,\theta)\right]_{r_0} + h_{33}^{\rm H}(r_0, \theta) = 0~,
\\
& \left[ \Psi_{\rm P}(r,\theta)\right]_{r_0} + \Psi_{\rm H}(r_0, \theta) = 0
\end{split}
\end{equation*}
to obtain $a_1$ and $b_1$.

\begin{figure}[htb]
\includegraphics[width=4cm]{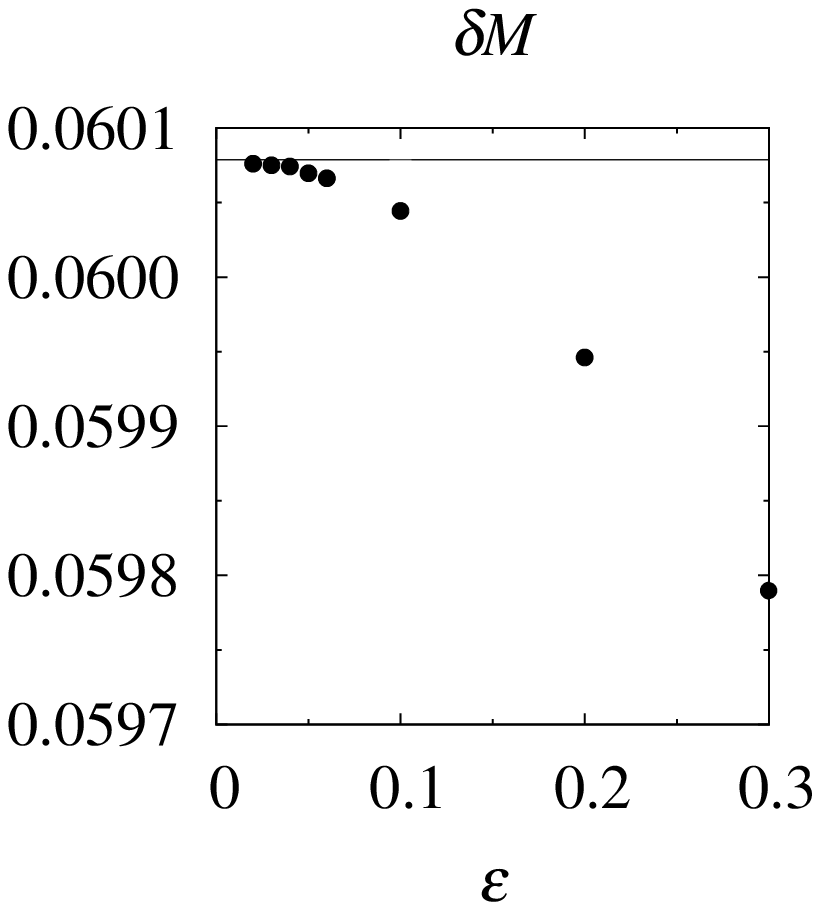}
\includegraphics[width=4cm]{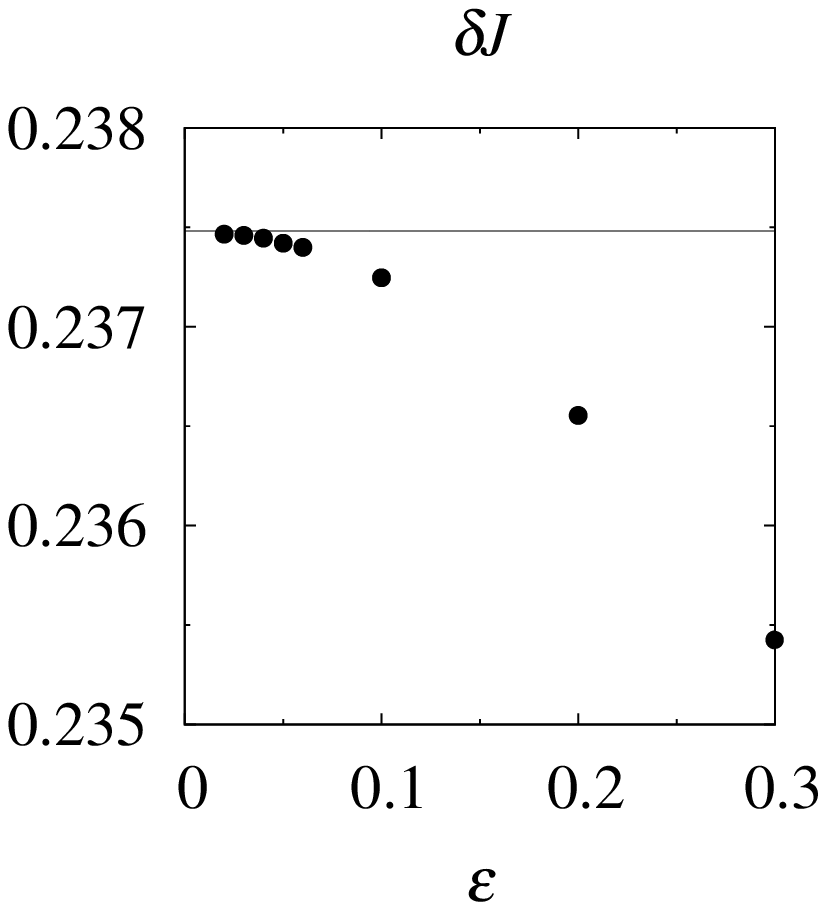}
\caption{
The plots of the numerically determined $\delta M$ and $\delta J$ \eref{deltaMdeltaJ}. As the accuracy of the fourth-order extrapolation is higher ($\epsilon\rightarrow 0$), $\delta M$ and $\delta J$ approaches to the analytic $M_{\rm ring}$ and $J_{\rm ring}$ (\eref{MJring}, the solid lines), respectively. 
}
\label{fig:epsilon}
\end{figure}

As a demonstration of the accuracy, we plot the numerically determined $\delta M$ and $\delta J$ \eref{deltaMdeltaJ}
as a function of $\epsilon$ in Fig. \ref{fig:epsilon}. 
Here, the meaning of $\epsilon$ is  as follows. 
When we evaluate the jump of, e.g.,  $\psi_1^{\rm P}(r,\theta)$ at $r=r_0$,
we evaluate $\psi_1^{\rm P}(r,\theta)$ up to $r=r_0\pm \epsilon$, 
and take the limit of $\epsilon\rightarrow 0$ by extrapolating 
$\psi_1^{\rm P}(r_0\pm\epsilon,\theta)$ to $\psi_1^{\rm P}(r_0,\theta)$ numerically 
by using the forth order polynomial.
If we use smaller $\epsilon$, it is expected that the accuracy of the result is improved. 
Thus, $\epsilon$ can be regarded as a parameter which controls the accuracy of the numerical results.  
In Fig. \ref{fig:epsilon}, we find that as $\epsilon$ becomes small, 
$-A(3M\Re(b_1)+\Re(b_2))$ and $-A\Im(a_2)$ approach $M_{\rm ring}$ and $J_{\rm ring}$ 
in \eref{MJring} respectively. 
This fact is an another evidence of the correctness of the results.

%




\begin{thebibliography}{99}
\bibitem{ref:KFWtypo}
During this work, we noticed that in Ref. \cite{kfw}
the term $+3\gamma\pmb{D}\rho(\pmb{\delta} +4\beta)\Psi$ was missed from $\psi_3$ in Eq. (105).
{A derivation of this term is given in Appendix \ref{section:psi3}.}
\end{thebibliography}
\end{document}